\documentclass[11pt,a4paper]{article}
\pdfoutput=1
\usepackage{jheppub,bm,booktabs,multirow}
\allowdisplaybreaks
\usepackage{amsmath,amssymb}
\usepackage{slashed}
\usepackage{epsfig}
\usepackage{graphicx}
\usepackage{float}
\usepackage[utf8]{inputenc}
\usepackage{color}
\usepackage[abs]{overpic}
\usepackage{xcolor,varwidth}
\usepackage[normalem]{ulem}

\def\sumint{\sum\hspace{-1.3em}\int}

\preprint{\begin{flushright}
CERN-TH-2019-059 \\ 
MIT-CTP 5118
\end{flushright}}

\begin{document}
\title{Resummation of Boson-Jet Correlation at Hadron Colliders}
\author[a]{Yang-Ting Chien}
\author[b]{\!, Ding Yu Shao}
\author[b]{and Bin Wu}
\affiliation[a]{Center for Theoretical Physics Massachusetts Institute of Technology, Cambridge, MA 02139}
\affiliation[b]{CERN, Theoretical Physics Department, CH-1211, Geneva 23, Switzerland}

\emailAdd{ytchien@mit.edu}
\emailAdd{dingyu.shao@cern.ch}
\emailAdd{b.wu@cern.ch}


\abstract{We perform a precise calculation of the transverse momentum ($\vec{q}_T$) distribution of the boson+jet system in boson production events. The boson can be either a photon, $W$, $Z$ or Higgs boson with mass $m_V$, and $\vec{q}_T$ is the sum of the transverse momenta of the boson and the leading jet with magnitude $q_T=|\vec q_T|$. Using renormalization group techniques and soft-collinear effective theory, we resum logarithms  $\log(Q/q_T)$ and $\log R$ at next-to-leading logarithmic accuracy including the non-global logarithms, where $Q$ and $R$ are respectively the hard scattering energy and the radius of the jet.  Specifically, we investigate two scenarios of $p^J_T \lesssim m_V$ or $p^J_T \gtrsim m_V$ in $Z$+jet events, and we examine the $q_T$ distributions with different jet radii and study the effect of non-global logarithms. In the end we compare our theoretical calculations with Monte Carlo simulations and data from the LHC.
}
\maketitle

\section{Introduction}

The production of photons, $W$'s, $Z$'s and Higgs bosons are important processes which allow us to test the Standard Model and extract its fundamental parameters. With precise calculations of the cross sections, they also give opportunities to search for physics beyond the Standard Model through deviations from Standard Model predictions. At hadron colliders, initial state radiation caused by the strong interaction contributes and necessarily affects the boson distributions. Moreover, energetic jets can be produced in association with the boson production. Therefore the understanding of the boson distribution is always complicated by the presence of hadronic activities in the events, which are governed by quantum chromodynamics (QCD). Using fixed-order calculations in perturbative QCD one can systematically improve the description of hadronic radiation. However, in certain regimes the fixed-order perturbative expansion diverges so that an all-order resummation is necessary for the validity of theoretical predictions. This happens when characteristic energy scales relevant in the process become hierarchical so that large logarithms of scale ratios can spoil the validity of fixed-order calculations. The transverse momentum $p_T$ distribution of the lepton pair in the Drell-Yan process is a classic example which requires the resummation of $\log (M_V/p_T)$ in the regime of $p_T\ll m_V$, where $M_V$ is the vector boson mass. This can be achieved to all-orders using the standard formalism by Collins, Soper and Sterman (CSS) \cite{Collins:1984kg}. Alternatively, the logarithms can also be resummed using renormalization group~(RG) techniques and soft-collinear effective theory (SCET) \cite{Bauer:2000yr,Bauer:2001yt, Bauer:2002nz, Beneke:2002ph} (see \cite{Becher:2014oda} for a review) as discussed in \cite{Gao:2005iu,Mantry:2009qz,Becher:2010tm,Becher:2011xn,Neill:2015roa,Ebert:2016gcn,Scimemi:2016ffw,Kang:2017cjk,Ebert:2018gsn}. More recently, the resummation has been performed at next-to-next-to-next-to-leading logarithmic accuracy \cite{Becher:2012yn,Bizon:2017rah,Bizon:2018foh,Chen:2018pzu}. 

\begin{figure}[h]
\begin{center}
  \includegraphics[width=0.7\textwidth]{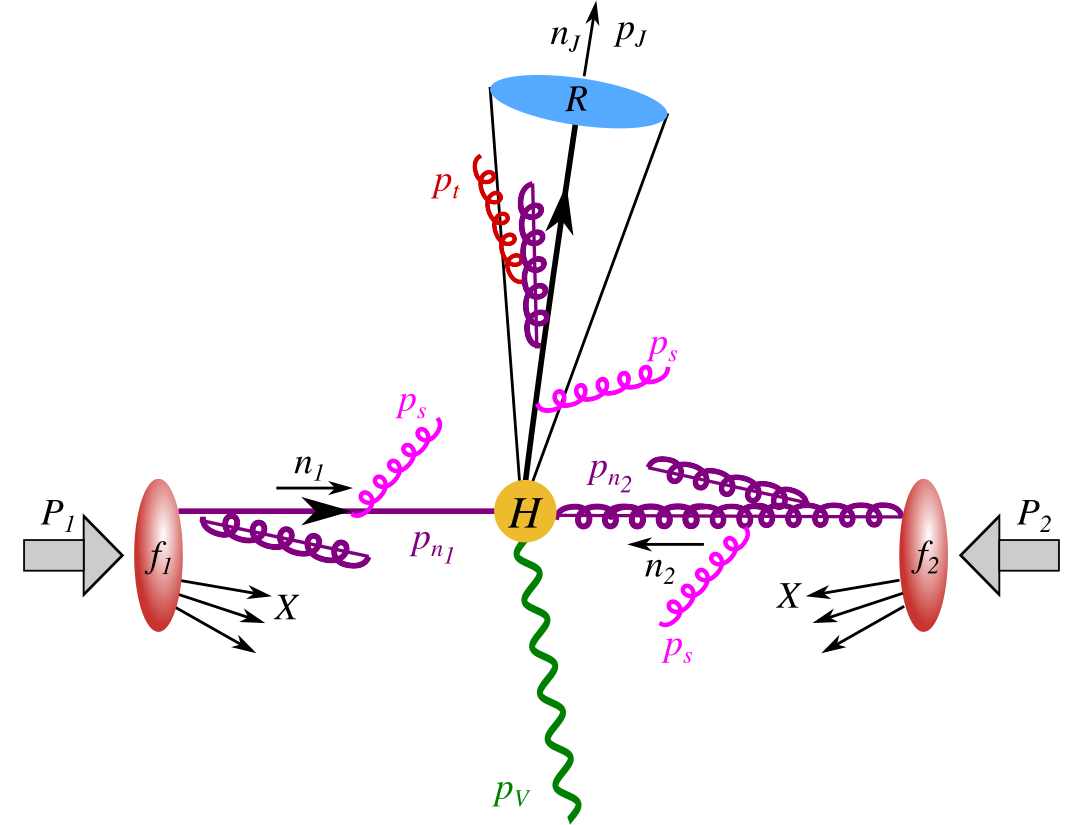}
\end{center}
  \caption{Boson+jet production in hadron collisions. Here $p_V$ and $p_J$ are the momenta of the color singlet boson and the jet, and $R$ is the jet radius.  By definition $\vec{q}_T=\vec{p}^J_{T}+\vec{p}^{\,V}_{T}$.  The modes relevant for the observable $q_T$ include the soft modes with momentum $p_s$, and the collinear modes along the two beam directions ($n_1$ and $n_2$) and the jet direction ($n_J$). Small-angle soft modes are taken as an independent degree of freedom from those emitted from the jet at wide angle, and its momentum is denoted as $p_t$. The $n_1$-collinear and $n_2$-collinear modes and soft modes all have a transverse momentum $\sim q_T$, while the $n_J$-collinear modes carry most of the jet momentum.}
\label{fig:eft}
\end{figure}

In this paper, we study the situation in which the boson has significant transverse momentum recoiling against hadronic activities consisting of jets in the final states. Specifically, we consider the $q_T$ distribution of the boson and the leading jet system where $\vec q_T$ is the vector sum of the transverse momenta of the two objects and $q_T=|\vec q_T|$, as illustrated in figure \ref{fig:eft}. In the $2\rightarrow 2$ scattering, boson+jet back-to-back limit the value of $q_T$ is zero, although the boson and the jet can have large transverse momenta. In the small $q_T$ regime, the soft and collinear emissions induce large logarithms of $\log (Q/q_T)$ which need to be resummed, where $Q$ represents the hard scattering energy. The situation is similar in di-jet production where $\vec q_T$ is defined as the vector sum of the transverse momenta of the two leading jets, and the resummation of $\log (Q/q_T)$ at next-to-leading logarithmic (NLL) accuracy without non-global logarithms (NGLs) \cite{Dasgupta:2001sh,Dasgupta:2002bw} was carried out using the CSS formalism in \cite{Sun:2014gfa,Sun:2015doa}. Similarly, the $\log (Q/q_T)$ resummation at NLL level was also performed for photon($\gamma$)+jet \cite{Chen:2018fqu}, $Z$+jet \cite{Sun:2018icb} and top quark+jet production \cite{Cao:2018ntd,Cao:2019uor}. More recently, the NLL resummation in $\gamma$+jet production was also carried out using SCET \cite{Buffing:2018ggv}.

The purpose of this paper is to derive an all-order expression in SCET for the systematic resummation of $\log(Q/q_T)$ in boson+jet production at small $q_T$, including the resummation of the associated jet radius logarithms $\log R$ as well as NGLs \footnote{Recently, much progress was made in the study of NGL resummation \cite{Hatta:2013iba,Caron-Huot:2015bja,Larkoski:2015zka,Becher:2015hka,Neill:2015nya,Becher:2016mmh,Becher:2016omr,Larkoski:2016zzc,Neill:2016stq,Becher:2017nof,Hatta:2017fwr,Martinez:2018ffw,Balsiger:2018ezi,Neill:2018mmj,Neill:2018yet,Balsiger:2019tne}.}. The precise understanding of this observable in proton-proton collisions then forms the baseline of such hard probes in nucleus-nucleus collisions where a hot and dense QCD medium called the quark-gluon plasma (QGP) is produced. Through interactions with the medium, jets in the event can be significantly modified while the color-singlet boson remains intact that can serve as a robust reference of the hard scattering process. This makes boson+jet production a useful channel for studying the properties of QGP though the relation between transverse momentum broadening and energy loss of jets in high-energy nuclear collisions \cite{Baier:1996sk}, which requires a proper resummation of large logarithms \cite{Mueller:2016gko, Mueller:2016xoc,Chen:2018fqu}. The kinematic information of the boson+jet system has been explored quite extensively \cite{Dai:2012am,Neufeld:2012df,Wang:2013cia,Casalderrey-Solana:2015vaa,Chien:2015hda,KunnawalkamElayavalli:2016ttl,Kang:2017xnc}. For example, the $q_T$, the boson-jet momentum imbalance $X_{JV}\equiv p_T^J/p_T^V$, and the azimuthal angle decorrelation $|\Delta \phi_{JV}|$: the azimuthal angle between the jet and the boson as measured along the beam direction, have been experimentally studied in $Z$+jet \cite{ Aad:2012en, Chatrchyan:2013tna,Khachatryan:2016crw,Sirunyan:2017jic, Sirunyan:2018cpw} and $\gamma$+jet \cite{Aaboud:2017kff} events at the LHC.

The rest of the paper is organized as follows. In section \ref{sec:factorization}, we analyze all the relevant degrees of freedom which contribute to $q_T$. We give a detailed derivation of our factorized expression  (\ref{eq:dsigmaFinal}) using a two-step matching procedure in SCET. In section \ref{sec:resum}, we discuss the renormalization of all the bare functions entering  (\ref{eq:dsigmaFinal}) and give an all-order resummation formula in  (\ref{eq:dsigmaResumFinal}). We explain the relation between our resummation formula with those in \cite{Chen:2018fqu,Sun:2018icb, Buffing:2018ggv}. The anomalous dimensions relevant for the NLL resummation are also given in this section. In section \ref{sec:LL} we analyze the Sudakov double logarithms, while in section \ref{sec:NLL} we perform the resummation of $\log(Q/q_T)$ at NLL accuracy for $Z$+jet production, including $\log R$ and NGL resummation. In section \ref{sec:summary} we summarize and discuss some intriguing issues for future studies. In appendices \ref{app:hard} and \ref{app:ad}, we list the tree-level amplitudes of partonic $V$+jet production and the anomalous dimensions used in this calculation. In appendix \ref{app:lo} we give the LO singular terms for $q_T$ distribution.

\section{Factorized Expression for $q_T$ in Boson+Jet Production}
\label{sec:factorization}

We derive a factorized expression of the differential cross section ${d\sigma}/{d^2q_T}$ for the process 
\begin{align}
N_1(P_1) + N_2(P_2) \to \text{boson}(p_V) + \text{jet}(p_J) + X,
\end{align}
where $X$ stands for all the produced particles in the event except for the boson and the particles of the leading jet. As defined previously, the observable $\vec{q}_T$ is the sum of the transverse momenta of the boson and the leading jet. The boson can be either the $W$, $Z$, $\gamma$ or the Higgs boson. We will first focus on the $q_T$ region where $\Lambda_{\rm QCD} \ll q_T\ll Q$ and $Q$ is a hard scattering energy scale depending on the leading-jet transverse momentum $p_T^J$ (and, for a massive boson, the boson mass $m_V$). We will discuss the factorization of the cross section in SCET and resum large logarithms using RG techniques. 

\subsection{Degrees of Freedom}

For $q_T\ll Q$, the dominant contributions to $q_T$ come from soft particles in all out-of-jet directions or collinear particles along the beam directions, with transverse momenta of the order $q_T$. As illustrated in figure \ref{fig:eft}, such radiation can be either soft with $p_s\sim q_T$, or collinear to the two beam directions $n_1$ and $n_2$ with $p^{n_1}_{T}\sim q_T$ or $p^{n_2}_{ T}\sim q_T$. In this paper all the calculations are carried out in the small $R$ limit. In this case, the small-angle soft mode along the jet direction can be singled out as an independent degree of freedom \cite{Becher:2015hka,Chien:2015cka,Becher:2016mmh,Kolodrubetz:2016dzb}. Such soft radiation is sensitive to the jet direction and the jet boundary and will be referred to as the coft mode in the following discussions, cf. \cite{Becher:2015hka,Becher:2016mmh}. While wide-angle, soft radiation is only sensitive to the total color charge of the jet, the coft mode can resolve any possible collinear constituents of the jet. If a coft radiation is emitted outside the jet it will contribute to the observable $q_T$. We need to consider the coft mode in order to account for multiple out-of-jet radiation and resum the potentially large logarithms of $\log R$. 

The above kinematic analysis shows that the relevant SCET degrees of freedom for the calculation of $q_T$ in this process include the following modes as illustrated in figure \ref{fig:eft}\footnote{
We do not include the Glauber mode which is responsible for the breakdown of the transverse-momentum factorization at higher orders as discussed in \cite{Collins:2007nk, Rogers:2010dm,Catani:2011st, Forshaw:2012bi}. 
Interested readers are referred to \cite{Rothstein:2016bsq} for a systematic study of the Glauber mode in SCET.},
\begin{align}\label{eq:scalings}
\text{$n_1$-{\rm collinear:}}~p_{n_1}^\mu &\sim  Q \,(\lambda^2,1,\lambda)_{n_1\bar n_1}, \notag \\
\text{$n_2$-{\rm collinear:}}~p_{n_2}^\mu &\sim  Q \,(\lambda^2,1,\lambda)_{n_2\bar n_2}, \notag \\
\text{$n_J$-{\rm collinear:}}~p_{n_J}^\mu &\sim  p^J_{T} \, (R^2,1,R)_{n_J \bar n_J}, \notag \\
{\rm soft:}~p_s^\mu &\sim  Q \,(\lambda, \lambda, \lambda), \notag \\
{\rm coft:}~p_{t}^\mu &\sim Q \lambda \, (R^2,1,R)_{n_J \bar n_J},
\end{align}
where $\lambda=q_T/Q$ is the power counting parameter. The auxiliary light-like vectors $\bar n_i$ satisfy $n_i\cdot \bar n_i=2$ for $i=1,2$ and $J$ and we choose $\bar n_1=n_2$ and $\bar n_2=n_1$. As we shall discuss in section \ref{sec:LL}, $Q$ is the hard scattering scale in the process and it may be parametrically different from $p^J_T$ if the boson is massive. Here all the momenta $p^\mu=(n_i\cdot p,\bar n_i\cdot p,\vec p_{n_i\perp})$ are expressed using light-cone coordinates with light-like vectors $n_i$ and $\bar n_i$, as denoted by the subscripts $n_i\bar n_i$ in (\ref{eq:scalings}). We denote the transverse momenta perpendicular to the two beam directions $n_1$ and $n_2$ by the subscript $_T$, and the transverse momenta perpendicular to $n_J$ by the subscript $_\perp$.

\subsection{Derivation of the Factorized Expression}

We derive the factorized expression using SCET with all the degrees of freedom in (\ref{eq:scalings}). The derivation is carried out in a two-step procedure similar to the one in \cite{Becher:2015hka, Becher:2016mmh}. 

\subsubsection{Matching of QCD onto an intermediate  SCET}
The intermediate SCET (more specifically, SCET$_\text{II}$ \cite{Bauer:2001yt}), 
includes $n_1$-, $n_2$- and $n_J$-collinear fields and one soft gluon field. In this step, the hard mode is integrated out and encoded in the Wilson coefficients $\mathcal{C}$. The local collinear gauge invariance demands that collinear fields along different directions do not directly interact with each other. That is, the hard and collinear modes factorize.

Let us denote collectively the infrared (soft or collinear) particles by $X_{\rm IR}$ and write the differential cross section for the process $N_1(P_1)+N_2(P_2)\to \text{boson}(p_V)+X_{\rm IR}$ as
\begin{align}\label{eq:dsigma}
\frac{d\sigma}{d{X_{\rm IR}} dy_V d^2 p^V_{T}}=\frac{1}{2s}\frac{1}{2(2\pi)^3} \int d^4 x e^{-i p_V\cdot x}
\langle P_1 P_2|\mathcal{H}^\dagger(x)|X_{\rm IR} \rangle \langle X_{\rm IR} |\mathcal{H}(0)|P_1 P_2\rangle,
\end{align}
where $|X_{\rm IR}\rangle $ is a product of $n_i$-collinear states $|X_{n_i}\rangle$ and soft state $|X_s\rangle$, $dX_{\rm IR}$ denotes the measure for the $n-$body relativistically invariant phase space of $X_{\rm IR}$ and $y_V$ is the boson rapidity. Generically, the leading-order operators are built out of three collinear fields along the three collinear directions of the beams and the jet, and the effective Hamiltonian $\mathcal{H}$ takes the form
\begin{align}\label{eq:H}
\mathcal{H}(x)=\int \{dt\}\mathcal{C}^{a_1a_2 a_J}_{\alpha_1\alpha_2 \alpha_J} (\epsilon, Q, \{t\})[\phi_{n_1}]^{\alpha_1}_{a_1}(x+t_1 \bar n_1)[\phi_{n_2}]^{\alpha_2}_{a_2}(x+t_2 \bar n_2) [\phi_{n_J}^\dagger]^{\alpha_J}_{a_J}(x+t_J \bar n_J),
\end{align} 
where $\{dt\}\equiv dt_1 dt_2 dt_J$ is the integration measure and $\mathcal{C}$ is the Wilson coefficient. The field $[\phi_{n_i}]^{\alpha_i}_{a_i}$ represents an $n_i$-collinear field carrying a color index $a_i$ and a Dirac or Lorentz index $\alpha_i$, and it can be either a collinear quark field or a collinear gluon field, which are given, respectively, by \cite{Hill:2002vw}
\begin{align}\label{eq:collinearFields}
\chi_{n_i}(x)=W_{n_i}^\dagger (x) \frac{\slashed{n}_i\slashed{\bar{n}}_i}{4} \psi_{n_i}(x),\qquad \mathcal{A}_{n_i\perp}^\mu=\frac{1}{g}W_{n_i}^+(x) \left[ i D_{n_i\perp}^\mu, W_{n_i}(x)\right],
\end{align}  
where the $n_i$-collinear covariant derivative is defined as $D_{n_i\perp}^\mu\equiv\partial^\mu_\perp +g A_{n_i\perp}^{\mu}$, and $W_{n_i}$ is the $n_i$-collinear Wilson line. The set of fields $\{\chi_{n_i}, \bar\chi_{n_i}, \mathcal{A}_{n_i\perp}^\mu\}$ form the building blocks to construct the effective operators. 

Since the $n_1$- and $n_2$-collinear modes are not directly measured and will go along the beams, we need to sum over these collinear states. Using the fact that initial colliding hadrons are color neutral and keeping only the leading contribution in $\bar n_i\cdot P_i$, we have 
\begin{align}\label{eq:beam}
  &\underset{X_{n_i}}{\sumint} 
  \langle P_i|[\phi_{n_i}^{f\dagger}]^{\alpha'_i}_{a'_i}(x+t'_i \bar n_i)|X_{n_i}\rangle \langle X_{n_i}|[\phi^f_{n_i}]^{\alpha_i}_{a_i}(t_i \bar n_i)|P_i\rangle \notag\\
  &\hspace{4cm}=\frac{\delta_{a'_i a_i}}{2d_i}\int_0^{1} \frac{d\xi_i}{\xi_i}~ {P}_{n_i}^{\alpha'_i \alpha_i} \mathcal{B}_{f/N_i}(\xi_i,x_T,\epsilon)e^{i \xi_i \bar n_i \cdot P_i\left( \frac{n_i\cdot x}{2}+t_i'-t_i \right)},
\end{align}
where $\epsilon=(4-d)/2$ in dimensional regularization and $i=1,2$ labeling the beam. The factor $d_i$ is the dimension of the color representation of the field $\phi_{n_i}^f$, and $P_{n_i}^{\alpha'_i \alpha_i}$ is the projector defined as follows\footnote{For gluon beam functions, another projector $\frac{x_T^{\alpha'_i} x_T^{\alpha_i}}{x_T^2}-\frac{g_T^{\alpha'_i\alpha_i}}{2}$ needs to be included in the study of, e.g., the Higgs $p_T$ distribution in the $gg\to H^0$ production channel \cite{Catani:2010pd, Becher:2011xn}. However, for the process studied in this paper, one can show that the contribution from this projector vanishes at NLL level. Hence, it is neglected here and in the following sections.},
\begin{align}
P_{n_i}^{\alpha'_i \alpha_i}=\left\{
\begin{array}{ll}
\frac{1}{2}\left(\slashed{n}_i\right)^{\alpha'_i\alpha_i} \xi_i \bar n_i \cdot P_i ~~~&\text{for quarks and antiquarks},\\
\frac{n_1^{\alpha_i} n_2^{\alpha'_i}+n_1^{\alpha'_i} n_2^{\alpha_i}}{2} -g^{\alpha'_i\alpha_i} \equiv -g_T^{\alpha'_i\alpha_i} ~~~&\text{for gluons}.
\end{array}
\right. 
\end{align}
The function $\mathcal{B}_{f/N_i}$ is the beam function of the parton species $f$ \cite{Collins:1981uk, Collins:1981uw, Stewart:2009yx} in the $x_T$ space, which is the Fourier transform of the transverse-momentum dependent (TMD) parton distribution functions (PDFs). 

Next, we sum over the $n_J$-collinear particles and perform multipole expansion so that the $n_J$-collinear fields only depend on $n_J\cdot x$. Assuming $m$ $n_J$-collinear partons in the jet, we have 
\begin{align}
p_J^\mu=\sum\limits_{i=1}^m p_{J_i}^\mu\qquad\text{with $p_{J_i}^\mu=p^{J_i}_{T}(\cosh\eta_{J_i},\sin\phi_{J_i},\cos\phi_{J_i},\sinh\eta_{J_i})$}
\end{align}
where the four-momentum of the $i$-th collinear particle in the jet $p_{J_i}^\mu$ is expressed in terms of the transverse momentum $p^{J_i}_{T}$, the azimuthal angle $\phi_{J_i}$ and the pseudo-rapidity $\eta_{J_i}$ of the particle. Also, the jet direction $n_J = (1,\sin\phi_J/\cosh\eta_J,\cos\phi_J/\cosh\eta_J,\tanh\eta_J)$ with $\eta_J$ and $\phi_J$ respectively  the rapidity and the azimuthal angle of the jet. For reasons that will become clear later, we also assume that there are some $n_J$-collinear particles radiated outside the jet with a total momentum $p_t^\text{out}$. Similar to the discussion of beam functions, a color-neutral jet function $\mathcal{J}^k$ with the virtuality $p_J^2$ and the parton species $k$ can be defined as
\begin{align}\label{eq:jetcoft}
  &\delta_{a'_J a_J} P_{n_J}^{\alpha'_J \alpha_J} \mathcal{J}^k(p_J^2,\vec x_T,\epsilon)
\equiv (2\pi)^{d-1} \\
 &  \underset{X_{n_J}}{\sumint} \int_0^\infty dp_J^2 \, 
    e^{\frac{i}{2} \bar n_J\cdot p^\text{out}_{t} \vec{n}_{JT}\cdot \vec{x}_T} \,  \delta^{(d)}\left(p_J- \sum\limits_{i=1}^m p_{J_i}\right)
  \langle 0 |[\phi^k_{n_J}]^{\alpha'_J}_{a'_J}\left(0\right)|X_{n_J}\rangle \langle X_{n_J} |[\phi_{n_J}^{k\dagger}]^{\alpha_J}_{a_J}\left(0\right)|0\rangle,  \notag
\end{align}
  where we have neglected the dependence of the jet function on $n_1\cdot x$ and $n_2\cdot x$, which will be justified in the next subsection. Likewise, $P_{n_J}^{\alpha'_J \alpha_J}$ is the projector defined as
\begin{align}
P_{n_J}^{\alpha'_J \alpha_J}=\left\{
\begin{array}{ll}
\frac{1}{2}\left(\slashed{n}_J\right)^{\alpha'_J\alpha_J} \bar n_J \cdot p_J~~~&\text{for quarks and antiquarks},\\
\frac{n_J^{\alpha_J} \bar n_J^{\alpha'_J}+n_J^{\alpha'_J}\bar n_J^{\alpha_J}}{2} -g^{\alpha'_J\alpha_J}\equiv -g_\perp^{\alpha'_J\alpha_J}~~~&\text{for gluons}.
\end{array}
\right. 
\end{align}
After decoupling the soft fields from \eqref{eq:H}, we will have the product of three soft Wilson lines. Summing over the states of soft gluons gives
\begin{align}\label{eq:soft}
\mathcal{S}^{\bar a_1 \bar a_2 \bar a_J}_{\bar a'_1 \bar a'_2 \bar a'_J}(\vec{x}_T,\epsilon)\equiv
\langle 0|\bar{T}[&(S^\dagger)^{\bar a_1' a_1}_{n_1}(\vec{x}_T)(S^\dagger)^{\bar a'_2 a_2}_{n_2}(\vec{x}_T)(S)^{a_J\bar a_J' }_{n_J}(\vec{x}_T)] \notag\\
&T[(S)^{ a_1 \bar a_1}_{n_1}(0)(S)^{ a_2 \bar a_2}_{n_2}(0)(S^\dagger)^{\bar a_J a_J  }_{n_J}(0)]|0\rangle.
\end{align} 
The color structure of the soft function is the same as the gauge transformation of the amplitude squared for the process. From the gauge invariance, one can show that the above matrix element is always proportional to the unit color matrix for the processes studied in this paper. That is,\begin{align}
\mathcal{S}&^{\bar a_1 \bar a_2 \bar a_J}_{\bar a'_1 \bar a'_2 \bar a'_J}(\vec{x}_T,\epsilon)\equiv\mathcal{S}(\vec{x}_T,\epsilon)\delta^{\bar a_1' \bar a_1} \delta^{\bar a'_2 \bar a_2} \delta^{\bar a'_J \bar a_J}.
\end{align}
Plugging (\ref{eq:beam}), (\ref{eq:jetcoft}) and (\ref{eq:soft}) into (\ref{eq:dsigma}) (with $X_{\rm IR}$ summed over as we have done), we have
\begin{align}\label{eq:dsigmastep1}
  \frac{d\sigma}{ d^2  p_{T}^J d^2 p_{T}^V d\eta_J dy_V }=\sum\limits_{ijk}\int \frac{d^2 x_T}{(2\pi)^2} e^{i\vec{q}_{T}\cdot \vec{x}_T} &\mathcal{S}_{ij\to Vk}(\vec{x}_T,\epsilon)\mathcal{B}_{i/N_1}(\xi_1,x_T,\epsilon)\mathcal{B}_{j/N_2}(\xi_2,x_T,\epsilon)\notag\\
  &\times\mathcal{H}_{ij\to V k}(\hat s, \hat t, m_V,\epsilon)\mathcal{J}^k(p_J^2,\vec{x}_T,\epsilon),
\end{align}
where the sum runs over all parton species $i,j,k = q, \bar q, g$. The hard function is identified as
\begin{align}
\label{eq:hard}
\mathcal{H}_{ij\to V k}\equiv &\frac{1}{16\pi^2 s^2} \frac{1}{\xi_1} \frac{1}{\xi_2}\left|\overline{M}\left(\xi_1 P_1, \xi_2 P_2\to p_J, p_V\right)\right|^2,
\end{align} 
with $\xi_{1}$ and $\xi_{2}$ completely determined by the conservation of the $+$ and $-$ components of the partonic momenta in the basis vectors $n_1$ and $\bar n_1$.

\subsubsection{Separating coft modes from $n_J$-collinear modes}

In this step, we match the purely collinear theory along the jet direction onto an effective theory where the collinear field is split into two submodes as
\begin{align}
\phi_{n_J} \to \phi_{n_J} + \phi_{t},
\end{align}
and, accordingly, 
\begin{align}
    |X_{n_J}\rangle\to|X_{n_J} X_t\rangle.
\end{align}
We distinguish genuine collinear momenta from the coft ones, and the corresponding momentum scalings are shown in \eqref{eq:scalings}.  Here the coft field describes low energy radiation which can resolve the substructure of the jet, and it is emitted from one of the collinear partons in the jet at an angle $\theta \lesssim R$. In effective theory languague one can take such coft radiation as being an independent mode \cite{Becher:2015hka, Becher:2016mmh}. In this effective theory the soft sector is a combination of soft radiation which can not resolve the detailed structure of the three collinear sectors, as well as coft radiation sourced by the collinear constituents of the jet\footnote{This can also be justified with color coherence \cite{Dokshitzer:1987nm}.}. In the limit $q_T\ll p_T^J$, the genuine $n_J$-collinear particles are kinematically forbidden to be radiated outside the jet while coft modes are allowed to be either inside or outside the jet. Inside the jet, the contribution from the coft modes to the jet momentum can be neglected. Hence, the $m$ $n_J$-collinear particles introduced in the previous subsection are genunine $n_J$-collinear particles while those outside the jet are coft particles with a total momentum $p_t^\text{out}$. 

The separation of the coft modes from the $n_J$-collinear modes modifies the jet function \eqref{eq:jetcoft} by organizing the coft radiation into the coft Wilson lines \cite{Becher:2015hka}. In the following discussion we use the notations adopted in \cite{Becher:2015hka, Becher:2016mmh} and write the amplitude squared for the $n_J$-collinear particles as
\begin{align}\label{eq:jetfun1}
\left|\mathcal{M}^k_m(p_J;\{ \underline{ p_J} \})\right\rangle\left\langle\mathcal{M}^{k\dagger}_m(p_J;\{ \underline{ p_J} \})\right| 
,
\end{align}
 where the compact notation $\{\underline{p_{J}}\}$ stands for the set of collinear parton momenta $p_{J_i}$. Then, the collinear Wilson line $W_{n_J}$ in the definition of the $n_J$-collinear field in (\ref{eq:collinearFields}) is replaced by
\begin{align}\label{eq:WU}
  W_{n_J}\to W_{n_J} U_{\bar{n}_J}
\end{align}
with the coft Wilson line $U_{\bar{n}_J}$ along the $\bar{n}_J$ direction which organizes coft radiation emitted from the $n_1$- and $n_2$-collinear directions in the small $R$ limit. Again, for brevity the collinear field $\phi_{n_J}$ dressed with coft radiation along the $\bar n_J$-direction due to the replacement in (\ref{eq:WU}) is written as $\phi_{n_J}\to \phi_{n_J} U_{\bar n_J}$. Also, each $n_J$-collinear parton is dressed with a coft Wilson line $U_{n_{J_i}}$ with $n_{J_i}=(1,\vec{p}_{J_i}/|\vec{p}_{J_i}|)$. This means that separating out the coft modes is equivalent to replacing $\left|\mathcal{M}_m(p_J;\{ \underline{ p_J} \})\right\rangle$ in \eqref{eq:jetfun1} with
\begin{align}
\left|\mathcal{M}_m(p_J;\{ \underline{ p_J} \})\right\rangle\rightarrow \bm{U}_{\bar n_J}(0)\prod\limits_{i=1}^m \bm{U}_{n_{J_i}}(0)\left|\mathcal{M}_m(p_J;\{ \underline{ p_J} \})\right\rangle.
\end{align}
Since $n_J\cdot p_{J}\ll \bar n_J\cdot p_{J}$, one has $\sqrt{p_J^2}\ll \bar n_J \cdot p_J$. Multipole expanding the integrand around $p_J^2=0$ in (\ref{eq:jetfun1}) gives
\begin{align}
  \int_0^\infty dp_J^2\,\delta(n_J\cdot p_J-\sum\limits_{i=1}^m n_J\cdot p_{J_i}) = \bar n_J\cdot p_J.
\end{align}
From the above two equations, one finally has
\begin{align}\label{eq:Jreplace}
  \mathcal{J}^k(p_J^2, \vec x_T,\epsilon)\to\sum\limits_{m=1}^\infty\langle \bm{\mathcal{J}}^k_m(\{\underline{n_{J}}\},R\, p_J,\epsilon)\otimes \bm{\mathcal{U}}^k_m(\{\underline{n_{J}}\}, R \, \vec{x}_T,\epsilon) \rangle 
\end{align}
where $\left\langle\cdots\right\rangle\equiv\frac{1}{d_J}\text{Tr}[\cdots]$ denotes the trace over all the color indices divided by the dimension of the color representation of $\phi^k_{n_J}$, and $\otimes$ is a short-hand notation for $\prod\limits_{i=1}^m\int d\Omega_{\vec{n}_{J_i}}/(4\pi) $ with $\Omega_{\vec{n}_{J_i}}$ the solid angle of $\vec{n}_{J_i}$ in $d$-dimension. The jet function $\bm{\mathcal{J}}^k_m$ with $m$ collinear particles is defined as
\begin{align}\label{eq:jet}
{P_{n_J}^{\alpha'_J \alpha_J}}&\bm{\mathcal{J}}^k_m(\{\underline{n_{J}}\},R\,p_J,\epsilon)\equiv 2\bar n_J\cdot p_J (2\pi)^{d-1} \sum_{\rm spins}\prod\limits_{i=1}^m \int\frac{d E_{J_i} E_{J_i}^{d-3}}{(2\pi)^{d-2}}\delta\Big(\bar n\cdot p_J-\sum\limits_{i=1}^m \bar n\cdot p_{J_i}\Big)  \notag\\
&\times\delta^{(d-2)}\Big(\sum\limits_{i=1}^m\vec{p}_{{J_i}\perp}\Big)\Theta_\text{in}(\{ \underline{ p_J} \})\left|\mathcal{M}^k_m(p_J;\{ \underline{ p_J} \})\right\rangle\left\langle\mathcal{M}^{k\dagger}_m(p_J;\{ \underline{ p_J} \})\right|,
\end{align}
and the coft function $\bm{\mathcal{U}}_m$ takes the form
\begin{align}\label{coftdef}
&\bm{\mathcal{U}}_m (\{\underline{n_{J}}\}, R\,\vec{x}_T, \epsilon) = \\
&~~\underset{X_t}{\sumint} e^{\frac{i}{2} p_t^{\rm out} \cdot \bar n_J \vec{n}_{JT} \cdot \vec{x}_T}   \langle 0 | \bm{U}_{\bar n_J}^\dagger(0) \bm{U}_ {n_{J_1}}^\dagger(0) \cdots \bm{U}_{ n_{J_m}}^\dagger(0)    | X_t \rangle \langle X_t | \bm{U}_{\bar n_J}(0) \bm{U}_ {n_{J_1}}(0) \cdots \bm{U}_{ n_{J_m}}(0)    | 0 \rangle. \notag
\end{align} 
The set of $n_J$-collinear particles is defined by the anti-$k_t$ algorithm \cite{Cacciari:2008gp} which is used in jet reconstruction. The phase space constraint imposed by the sequential clustering can be quite complicated. Alternatively, here we require the angle $\Delta R_{ij}$ between each pair of collinear particles be smaller than the jet radius $R$,
\begin{align}
\Delta R_{ij}\equiv \sqrt{(\phi_i-\phi_j)^2+(\eta_i-\eta_j)^2}<R\qquad \text{with $i<j:1,2,\cdots, m$}.
\end{align}
In the small $R$ limit, the above requirement is equivalent to imposing the following step functions,
\begin{align}
\Theta_{\text{in}}(p_{J_i},p_{J_j})\equiv \theta\left(R^2-\frac{2 p_{J_i}\cdot p_{J_j}}{p^{J_i}_{T} p^{J_j}_{T}}\right),
\end{align}
which collectively is denoted by $\Theta_{\text{in}}(\{ \underline{ p_J} \})$. 
The jet algorithm constraint for a coft gluon with momentum $p_t$ is then equivalent to a cone jet algorithm since collinear particles are clustered and define the jet direction $n_J$,
\begin{align}\label{eq:coftout}
\Theta_\text{out}(p_t)\equiv 1-\Theta_\text{in}(p_t, n_J)=\theta\left[ \frac{n_J\cdot p_t}{\bar n_J\cdot p_t} - \left(\frac{R}{2\cosh \eta_J}\right)^2\right].
\end{align}
By making the replacement in \eqref{eq:Jreplace}, (\ref{eq:dsigmastep1}) then gives the final factorized expression 
\begin{align}\label{eq:dsigmaFinal}
 & \frac{d\sigma}{d^2q_{T} d^2 p_{T} d\eta_J dy_V }=\sum\limits_{ijk}\int \frac{d^2 x_T}{(2\pi)^2} e^{i\vec{q}_{T}\cdot \vec{x}_T}\mathcal{S}_{ij\to V k}(\vec{x}_T,\epsilon) \mathcal{B}_{i/N_1}(\xi_1,x_T,\epsilon)\mathcal{B}_{j/N_2}(\xi_2,x_T,\epsilon)\notag\\
  &\hspace{2cm}\times\mathcal{H}_{ij\to V k}(\hat s, \hat t, m_V,\epsilon)\sum\limits_{m=1}^\infty\langle \bm{\mathcal{J}}^k_m(\{\underline{n_{J}}\},R\, p_J,\epsilon)\otimes \bm{\mathcal{U}}^k_m(\{\underline{n_{J}}\}, R \, \vec{x}_T,\epsilon) \rangle.
\end{align}

\section{Resummation of Large Logarithms}\label{sec:resum}

In this section, we discuss the renormalization of the bare functions in  (\ref{eq:dsigmaFinal}) and the resummation of large logarithms by solving the corresponding RG equations. We also calculate the anomalous dimensions relevant for the resummation at NLL level. 

\subsection{Renormalization and Resummation}

The cross section is finite in the limit $\epsilon\to 0$ but all the bare functions in (\ref{eq:dsigmaFinal}) are divergent. In this paper, these functions are renormalized in the $\overline{\text{MS}}$ scheme. The divergent pieces of the bare functions are removed by the renormalization constants, and the anomalous dimensions can be calculated from them according to (\ref{eq:Gamma}). Then the resummation of large logarithms can be achieved by solving the RG equations. 

\subsubsection{Hard function}

The Wilson coefficient $\mathcal{C}$ in (\ref{eq:H}) is determined order-by-order in perturbation theory by a matching calculation in QCD and in SCET. In dimensional regularization, the ultraviolet (UV) divergence in the Wilson coefficient is identical to the infrared (IR) divergence in the corresponding on-shell amplitudes in perturbative QCD. Hence, the singularities in the hard function can be subtracted by a multiplicative renormalization constant $Z^{\rm H}_{ij\to Vk}$. From $Z^{\rm H}_{ij\to Vk}$ one can calculate the anomalous dimensions of the hard functions and resum large logarithms in $\mu_h/\mu$ by solving the RG equation
\begin{align}
  \frac{d}{d\log\mu}\mathcal{H}_{ij\to V k}(\hat s, \hat t, m_V,\mu)=\Gamma^{{\rm H}_{ij\to V k}}(\hat{s},\hat{t},m_V,\mu)\mathcal{H}_{ij\to V k}(\hat s, \hat t, m_V,\mu),
\end{align}
with the initial condition $\mathcal{H}_{ij\to V k}(\hat s, \hat t, m_V,\mu_h)$ at the hard scale $\mu_h\sim Q$ calculated in a matching calculation. 

\subsubsection{Soft function, beam function and collinear anomaly}

The calculation of soft and beam functions involves extra complication which is not seen in the calculation of the hard function.  Singularities unregularized by dimensional regularization arise in the calculation of these functions. However, such divergences are artificial because the product of the soft and beam functions is in fact finite, which is a result independent of the regulator. In this paper, we regularize such divergences by modifying the phase-space integrals as \cite{Becher:2011dz}
\begin{align}\label{eq:rapidityreg}
\int d^dk \to \int d^d k \left(\frac{\nu}{n_1\cdot k}\right)^\alpha.
\end{align}
Note that we have chosen the common factor of $n_1\cdot k$ in the regulator. Moreover, the scale separation in (\ref{eq:scalings}) is broken due to loop corrections, and the hard scale shows up in the perturbative calculation of soft and beam functions. This is referred to as the collinear anomaly by the authors of \cite{Becher:2010tm} \footnote{Alternatively, the collinear anomaly can be dealt with using the rapidity RG method \cite{Chiu:2011qc,Chiu:2012ir}. }. By refactorizing out the collinear anomaly, the product of beam and soft function can be written as \cite{Becher:2015gsa}
\begin{align}\label{coll}
&\mathcal{B}_{i/N_1}(\xi_1, x_T, \mu ) \mathcal{B}_{j/N_2}(\xi_2, x_T, \mu )   \mathcal{S}_{ij\to Vk}(\vec x_T, \mu) = \notag \\ 
&\hspace{2cm} \left( \frac{x_T^2 \, \hat s}{b_0^2} \right)^{-(C_i + C_j)F_\perp(x_T,\mu)} B_{i/N_1}(\xi_1, x_T, \mu ) B_{j/N_2}(\xi_2, x_T, \mu )   S_{ij\to Vk}(\vec x_T, \mu), 
\end{align}
where the hard scale dependence is factored out with the exponent $F_{\perp}$ only depending on $x_T$ and the scale $\mu$. Here $b_0=2 e^{-\gamma_E}$ and $C_i$ is the Casimir operator of the color representation of the parton $i$.  The divergence in the product of the soft and beam functions on the l.h.s. of \eqref{coll} is to be removed by an overall multiplicative renormalization factor, denoted by $Z^{\mathcal {SBB}}(\vec x_T, \mu,\epsilon)$. In order to construct a universal definition of the beam function (at least in the boson+jet processes considered in this paper), we take $Z^{\mathcal {SBB}}$ as a product of the renormalization constants of the collinear anomaly $Z^{\rm CA}$, the soft function $Z^{\rm S}$, and the beam functions $Z^{\rm B}_{i/N_1}$ and $Z^{\rm B}_{j/N_2}$,
\begin{align}
	Z^{\mathcal {SBB}}_{ij\to Vk} = Z^{\rm CA}_{ij} ~ Z^{\rm S}_{ij\to Vk} ~ Z^{\rm B}_{i/N_1} ~ Z^{\rm B}_{j/N_2}.
\end{align}
From these renormalization constants one can calculate the corresponding anomalous dimensions. The collinear anomaly exponent function $F_\perp(x_T, \mu)$, beam and soft functions $B_{f/N}$ and $S_{ij\to Vk}$ satisfy the following RG equations, respectively
\begin{align}
&\frac{d}{d\log\mu}F_\perp(x_T, \mu)=\gamma_\text{cusp}(\alpha_s),  ~~~~~\frac{d}{d\log\mu}B_{f/N}(\xi, x_T,\mu)=\Gamma^{{\rm B}_f}(\alpha_s) B_{f/N}(\xi, x_T,\mu), \notag\\
&\frac{d}{d\log\mu}S_{ij\to Vk}(\vec x_T,\mu)=\Gamma^{{\rm S}_{ij\to Vk}}(\alpha_s)S_{ij\to Vk}(\vec x_T,\mu).
\end{align}

\subsubsection{Jet function, coft function and non-global logarithms}

The calculations of jet and coft functions contain NGLs because of the restricted phase space due to jet definition. As discussed in \cite{Becher:2015hka, Becher:2016mmh}, the RG running of the jet and coft functions in the factorized expression (\ref{eq:dsigmaFinal}) automatically resums both global and non-global logarithms.

In the definition of the jet function in (\ref{eq:jet}), the energy of the $n_J$-collinear constituents are integrated over, which results in additional singularities. However, such singularities can be cancelled by the jet functions with lower parton multiplicity. Therefore, in general the renormalization constant of jet functions is a matrix \cite{Becher:2016mmh}, which is defined as
\begin{align}\label{eq:Jbare}
  \bm{\mathcal{ J}}_m(\{\underline{n}\},R\, p_J, \epsilon)= \sum\limits_{l=1}^m\bm{\mathcal{J}}_l(\{\underline{n}\},R\, p_J, \mu) \bm{Z}^{\rm J}_{lm}(\{\underline{n}\},\mu,\epsilon).
\end{align}
Similarly, the renormalized coft function is written as 
\begin{align}\label{eq:Ubare}
\bm{\mathcal{U}}_l(\{\underline{n}\},R\, \vec{x}_T, \mu)=\sum\limits_{m=l}^\infty \bm{Z}^{\rm U}_{lm}(\{\underline{n}\},R\, \vec{x}_T ,\mu,\epsilon)\hat{\otimes} \,\bm{\mathcal{U}}_m(\{\underline{n}\},R\, \vec{x}_T,\epsilon),
\end{align}
where $\hat{\otimes}$ denotes the integration over the $(m-l)$ additional directions. Note that $\bm{Z}^{\rm U}_{lm}$ is defined in a reversed way as opposed to the other renormalization constants. By the RG invariance of the physical cross section, the renormalization matrix of the coft function satisfies
\begin{align}\label{eq:ZU}
\bm{Z}^{\rm U}_{lm}=Z^{\rm H} Z^{\mathcal{SBB}}\bm{Z}^{\rm J}_{lm}.
\end{align}
In \cite{Becher:2016mmh,Becher:2016omr} one of the authors has explicitly verified that this matrix satisfies a renormalization group equation at two-loop level for non-global jet observables in electron-positron collisions. 

For the coft function we specifically extract the global renormalization constant  $Z^{\rm U}$ which removes the divergence in the coft function $\bm{\mathcal{U}}_1$,
\begin{align}\label{eq:ZU1}
\bm{\mathcal{U}}_1(\{\underline{n}\},R\, \vec{x}_T,\epsilon)=\bm{\mathcal{U}}_1(\{\underline{n}\},R\, \vec{x}_T,\mu) Z^{\rm U}. 
\end{align}
Accordingly, we define the non-global renormalization constant as
\begin{align}\label{eq:ZU_NG}
\bm{\hat Z}_{lm}\equiv \bm{Z}^{\rm U}_{lm}Z^{\rm U}
\end{align}
by separating out the global contribution. From \eqref{eq:ZU}, $\bm{Z}^{\rm J}_{lm}$ can hence be expressed as the product of non-global and global renormalization constants
\begin{align}\label{eq:ZJ}
\bm{Z}^{\rm J}_{lm} = \bm{\hat Z}_{lm}  (Z^{\rm U} Z^{\rm H} Z^{\mathcal{SBB}})^{-1}  \equiv \bm{\hat Z}_{lm}Z^{\rm J},
\end{align}
where we also introduce a global renormalization constant $Z^{\rm J}$ for the jet function. The evolution equations that resum both the global and non-global logarithms in the jet and coft functions can be obtained from (\ref{eq:Jbare}) and (\ref{eq:Ubare}). Differentiating both sides of these equations gives
\begin{align}\label{eq:JUrg}
\frac{d}{d\log\mu}\bm{\mathcal{J}}_m(\{\underline{n}\},\mu)=& \sum_{l=1}^m \bm{\mathcal{J}}_l(\{\underline{n}\},\mu)\underbrace{\left[\Gamma^{\rm J} \delta_{lm} \bm{1} - \bm{\hat{\Gamma}}_{lm}(\{\underline{n}\},\mu)\right]}_{ \bm{\Gamma}^{\rm J} }, \notag\\
\frac{d}{d\log\mu}\bm{\mathcal{U}}_l(\{\underline{n}\},\mu)=&\sum_{m=l}^\infty\underbrace{\left[\Gamma^{\rm U} \delta_{lm} \bm{1} + \bm{\hat{\Gamma}}_{lm}(\{\underline{n}\},\mu)\right]}_{\bm{\Gamma}^{\rm U}}\hat \otimes\,\bm{\mathcal{U}}_m(\{\underline{n}\},\mu),
\end{align}
where the diagonal entry represents the global anomalous dimensions $\Gamma^{\rm J}$ and $\Gamma^{\rm U}$, which can be calculated from $Z^{\rm J}$ and $Z^{\rm U}$ according to (\ref{eq:Gamma}).

\subsubsection{Resummed expression}\label{sec:ResumFinal}

Using the RG equations we can evolve each function from its characteristic scale where there are no large logarithms, and we get the following resummed expression 
\begin{align}\label{eq:dsigmaResumFinal}
&\frac{d\sigma}{d^2q_{T} d^2 p_{T} d\eta_J dy_V}=\sum\limits_{ijk}\int \frac{d^2 x_T}{(2\pi)^2} e^{i\vec{q}_{T}\cdot \vec{x}_T}e^{\int_{\mu_h}^\mu \frac{d\bar \mu}{\bar \mu} \Gamma^{{\rm H}_{ij\to V k}}(\bar \mu)}\mathcal{H}_{ij\to V k}(\hat s,\hat t,m_V,\mu_h)\notag\\
&\times \left( \frac{x_T^2 \, \hat s}{b_0^2} \right)^{-(C_i + C_j){F_\perp(\mu)}}e^{\int_{\mu_b}^\mu \frac{d\bar \mu}{\bar \mu} \Gamma^{{\rm W}_{ij\to V k}}(\bar \mu)} {S}_{ij\to V k}(\vec{x}_T,\mu_b) B_{i/N_1}(\xi_1,x_T,\mu_b) {B}_{j/N_2}(\xi_2,x_T,\mu_b) \notag\\
&\times  e^{\int_{\mu_t}^\mu \frac{d\bar \mu}{\bar \mu} \Gamma^{{\rm U}_{k}}(\bar \mu)+\int_{\mu_j}^\mu \frac{d\bar \mu}{\bar \mu} \Gamma^{{\rm J}_{k}}(\bar \mu)}U_{\rm NG}^k(\mu_t,\mu_j),
\end{align}
where $\Gamma^{{\rm W}_{ij\to V k}}\equiv\Gamma^{{\rm B}_i}+\Gamma^{{\rm B}_j}+\Gamma^{{\rm S}_{ij\to Vk}}$. The function $U_{\rm NG}^k$ includes NGL resummation, which is defined as 
\begin{align}\label{eq:UNG}
U_{\rm NG}(\mu_t,\mu_j)\equiv\sum_{l=1}^\infty \big\langle \bm{\mathcal{J}}_l(\{\underline{n}'\},R\,p_T,\mu_j) 
    \otimes \sum_{m\geq l}^\infty \bm{U}_{lm}(\{\underline{n}\},\mu_t,\mu_j)\,\hat{\otimes}\, \,
    \bm{\mathcal{U}}_m(\{\underline{n}\},R\,\vec x_T,\mu_t) \big\rangle 
\end{align}
with $\bm{U}(\{\underline{n}\},\mu_t,\mu_j) = {\rm \bf P} \exp\big[ \int_{\mu_t}^{\mu_j} d\log\mu \, \bm{\hat \Gamma}( \{\underline{n}\} ,\mu) \big]$, where ${\rm \bf P}$ denotes the path ordering in $\log \mu$. This evolution matrix generates additional collinear partons with $m\geq l$, therefore we define $\{\underline{n}'\}= \{n_1 , \dots , n_l\} $ and $\{\underline{n}\}=\{n_1 , \dots , n_l, n_{l+1}, \dots, n_m\}$ to distinguish these two configurations.   According to the momentum scalings in (\ref{eq:scalings}), one should choose the hard scale $\mu_h$, the soft and beam scale $\mu_b$, the jet scale $\mu_j$ and the coft scale $\mu_t$ with the following typical values
\begin{align}
	\mu_h\sim Q, ~~~~~~~ \mu_b\sim b_0/x_T, ~~~~~~~ \mu_j\sim R~p_T, ~~~~~~~ \mu_t\sim R \, b_0/x_T,
\end{align}
so that there are no residual large logarithms in the corresponding functions at these scales\footnote{As shown in section \ref{sec:LL}, the hard function can have additional logarithms because it depends on two scales $p_T^J$ and $m_V$.}.

\subsection{Anomalous Dimensions for NLL resummation}\label{subsec:adoneloop}
To perform NLL resummation, one needs to include tree-level hard, jet, beam, soft, and coft functions, and evolves them using two-loop cusp anomolous dimension and one-loop regular anomolous dimensions. In this section we will provide all the regular one-loop anomalous dimensions relevant for the NLL resummation. In the calculation we neglect the difference between $p_T$ and $p_T^J$ (recall that $\vec{p}_T\equiv(\vec{p}^J_{T}-\vec{p}^{\,V}_{T})/2$).

\subsubsection{Anomalous dimensions of hard, soft and beam functions}

The one-loop hard anomalous dimension is given by \cite{Becher:2011fc}
\begin{align}
  \Gamma^{{\rm H}_{ij\to V k}}&=\gamma_{\text{cusp}}(\alpha_s)\left[C_i\log\left(\frac{\hat{u}^2}{p_T^2 \mu^2}\right)+C_j\log\left(\frac{\hat{t}^{\,2}}{p_T^2 \mu^2}\right)+C_k\log\left(\frac{p_T^2}{\mu^2}\right)\right]+\gamma^{{\rm H}_{ij\to Vk}},
\end{align}
with
\begin{align}
  \gamma^{{\rm H}_{ij\to Vk}}\equiv 2\gamma^i(\alpha_s)+2\gamma^j(\alpha_s)+2\gamma^k(\alpha_s),
\end{align}
where $\gamma_\text{cusp}$ is the cusp anomalous dimension,  and $\gamma^f$ is the anomalous dimension of the parton species $f$ (see appendix \ref{app:ad}). In dimensional regularization, at one-loop only real emission diagrams contribute to the soft function. Using the covariant gauge one has
\begin{align}\label{eq:S}
S_{ij\to V k} (\vec{x}_\perp,\epsilon)  =&\, g_s^2 {\tilde \mu }^{2\epsilon}  \int \frac{d^d k }{(2\pi)^{d-1}} \left(\frac{\nu}{n_1 \cdot k}\right)^\alpha \delta^+ (k^2) e^{i k_T \cdot x_T} \left[ \left( C_i+C_j -C_k\right) \frac{n_1 \cdot n_2}{n_1 \cdot k \,k \cdot n_2} \right.\notag \\
&\left.+\left( C_i+C_k-C_j \right) \frac{n_1 \cdot n_J}{n_1 \cdot k \, k\cdot n_J}  + \left( C_j+C_k-C_i \right)\frac{n_2 \cdot n_J}{n_2 \cdot k \, k\cdot n_J} \right], 
\end{align}
with $\delta^+(k^2) = \delta(k^2)\theta(k^0)$ and $\tilde\mu^2\equiv\frac{\mu^2 e^{\gamma_E}}{4\pi}$. The evaluation of the soft function boils down to the calculation of the following master integrals,
\begin{align}
\omega_{ab} = g_s^2 {\tilde \mu }^{2\epsilon} \int \frac{d^d k }{(2\pi)^d} \left(\frac{\nu}{k^+}\right)^\alpha (2\pi)\delta^+ (k^2) e^{i k_T \cdot x_T} \frac{n_a \cdot n_b}{n_a \cdot k \, k\cdot n_b}.  
\end{align}
With the regulator we use, $\omega_{12}$ involves a scaleless integral and hence vanishes. The divergent parts of the other two integrals are given by 
\begin{align}\label{eq:omegaab}
\omega_{1J} & = \frac{\alpha_s}{4\pi} e^{(\epsilon+\alpha/2)L_\perp} \left(\frac{\nu}{\mu}\right)^\alpha \left[ \frac{2}{\alpha\,\epsilon} + \frac{2}{\epsilon} \Big[ \eta_J +  \log (-2i\cos\phi_x)\Big]  \right], \notag \\
\omega_{2J} & = \frac{\alpha_s}{4\pi} e^{(\epsilon+\alpha/2)L_\perp} \left(\frac{\nu}{\mu}\right)^\alpha \left[ \frac{2}{\epsilon^2} - \frac{2}{\alpha\,\epsilon} - \frac{2}{\epsilon} \Big[   \eta_J -  \log(-2i \cos\phi_x) \Big]  \right],
\end{align}
where $L_\perp\equiv\log\left(\frac{x_T^2\mu^2}{b_0^2}\right)$ and $\phi_x$ represents the azimuthal angle between  $\vec{x}_T$ and $\vec{n}_{JT}$. From the above expressions we can identify the soft anomalous dimension. 

For $1/x_T\sim q_T\gg \Lambda_{\rm QCD}$, one can calculate the beam functions from PDFs by an operator-product expansion \cite{Collins:1981uk, Collins:1981uw, Collins:1984kg}
\begin{align}
\mathcal{B}_{i/N}(\xi,x_T,\mu)= \sum_{j} \int_\xi^1 \frac{dz}{z}\mathcal{I}_{i\leftarrow j}(z,x_T,\mu) f_{j/N}(\xi/z,\mu).
\end{align}
If one chooses $\mu=\mu_{b}\equiv\frac{b_0}{x_T}$, the logarithms $\ln(x_T\mu)$ in $\mathcal{I}$ vanish. Since we are only resumming large logarithms of $\ln(\mu/\mu_{b})$ at NLL level in this paper, we will neglect the non-logarithmic terms in $\mathcal{I}$ at $O(\alpha_s)$ in the following sections and only need the anomalous dimensions of the beam functions.

Let us focus on the non-PDF anomalous dimensions. The divergent pieces of the bare beam functions using the rapidity regulator in (\ref{eq:rapidityreg}) take the form 
\begin{align}\label{eq:B}
&\mathcal{B}_{i/N_1}(\xi_1, x_T,\epsilon)=\frac{\alpha_s}{4\pi}\left[4 C_i e^{(\epsilon+\alpha) L_\perp} \left(\frac{\nu}{\mu}\right)^\alpha \left(\frac{\xi_1\bar n_1\cdot P_1}{\mu}\right)^{\alpha} \left(\frac{1}{\epsilon^2}-\frac{1}{\epsilon \alpha}\right)-\frac{\gamma^i_0}{\epsilon}\right]f_{i/N_1}(\xi_1, \mu)+\cdots,\notag\\
&\mathcal{B}_{j/N_2}(\xi_2, x_T,\epsilon)=\frac{\alpha_s}{4\pi}\left[ 4C_j e^{\epsilon L_\perp} \left(\frac{\nu}{\mu}\right)^\alpha \left(\frac{\xi_2\bar n_2\cdot P_2}{\mu}\right)^{-\alpha} \frac{1}{\epsilon \alpha}-\frac{\gamma^j_0}{\epsilon}\right] f_{j/N_2}(\xi_2, \mu)+\cdots.
\end{align}
From (\ref{eq:S}), (\ref{eq:omegaab}) and (\ref{eq:B}) one can easily verify the cancellation of all the $\alpha$-dependent terms in the soft and beam functions. 

Since the soft and beam functions have the same characteristic momentum scale $\sim 1/x_T$, one can evolve
the product of these functions from $\mu_{b}\sim 1/x_T$ to $\mu$ instead of running each one of them individually. We write
\begin{align}
\mathcal{B}_{i/N_1}(\xi_1, x_T, \mu ) & \mathcal{B}_{j/N_2}(\xi_2, x_T, \mu )   \mathcal{S}_{ij\to V k}(\vec{x}_T, \mu)\notag\\
 &\hspace{3.5cm} = \left( \frac{x_T^2 \, \hat s}{b_0^2} \right)^{-(C_i + C_j){F_\perp}(x_\perp,\mu)} \mathcal{W}_{ij\to Vk}(\vec{x}_T,\mu),
\end{align}
where the function $\mathcal{W}_{ij\to V k}$ satisfies the following evolution equations,
\begin{align}
\frac{d}{d\log\mu} \mathcal{W}_{ij \to Vk} = \underbrace{\left[ \left(C_i + C_j+C_k\right) \gamma_{\rm cusp}(\alpha_s) \log \left(\frac{x_T^2 \mu^2}{b_0^2} \right)+ \gamma^{{\rm W}_{ij\to Vk}}(\alpha_s) \right]}_{\Gamma^{{\rm W}_{ij\to Vk}}} \mathcal{W}_{ij\to V k}
\end{align}
with the anomalous dimension at one-loop level
\begin{align}
\gamma^{{\rm W}_{{ij\to Vk}}}_0=
8\, C_k\log(-2i \cos\phi_x)-2\gamma^i_0-2\gamma^j_0-\gamma^{\text{cusp}}_0\left[ C_i \log\left(\frac{\hat{u}^2}{\hat{s} p_T^2}\right)+C_j \log\left(\frac{\hat{t}^2}{\hat{s}p_T^2}\right)\right].
\end{align}

\subsubsection{Anomalous dimensions of jet and coft functions}

The global coft anomalous dimension $\Gamma^{{\rm U}}$ can be derived from the one-loop calculation of the coft function $\bm{\mathcal{ U}}_1$. Explicitly, $\bm{\mathcal{ U}}_1$ contains two Wilson lines, one along the $n_J$ direction and the other one along the $\bar n_J$ direction. After expanding Wilson lines in (\ref{coftdef}), at one-loop we have
\begin{align}
\bm{\mathcal{ U}}^k_1 (\vec{x}_T,\epsilon)  = & 2 {{ C}}_k\, g_s^2 \, {\tilde \mu }^{2\epsilon}  \int \frac{d^d p_{t} }{(2\pi)^d}  (2\pi)\delta^+ (p_{t}^2) e^{- \frac{i}{2} (\bar n_J\cdot p_{t})  n_J \cdot  x_T}  \frac{2}{n_{J} \cdot p_t \, p_t\cdot \bar n_J} \Theta_\text{out}(p_t).
\end{align}
We only need the divergent terms in order to obtain the anomalous dimension, and we find
\begin{align}\label{eq:ZUoneloop}
Z^{\rm U} \equiv 1 + \frac{\alpha_s}{4\pi} \Big\{ -\frac{2}{\epsilon^2}C_k   -\frac{1}{\epsilon}2 C_k  \left[L_\perp+2 \log\left(\frac{-2i\cos\phi_x}{R}\right) \right]\Big\},
\end{align}
which gives,
\begin{align}
  \Gamma^{{\rm U}_k}=
  C_k \gamma_{\text{cusp}} \log\left(\frac{R^2 b_0^2}{\mu^2 x_T^2}\right)+\gamma^{{\rm U}_k},~~~~~~~~{\rm with~}
  \gamma^{{\rm U}_k}_0\equiv-8 C_k \log(-2 i \cos\phi_x).
\end{align}
From the definition \eqref{eq:ZJ}, the global jet renormalization constant at one-loop is written as
\begin{align}
Z^{\rm J}\equiv
1 + \frac{\alpha_s}{4\pi}\left\{ \frac{2}{\epsilon^2} C_k + \frac{1}{\epsilon} \left[ 2C_k\log\left(\frac{\mu^2}{p_T^2R^2}\right) -\gamma^k_0\right]\right\},
\end{align}
and the anomalous dimension $\Gamma^{\rm J}_k$ has the form
\begin{align}\label{eq:GammakJ}
  &\Gamma^{{\rm J}_k}
  =-C_k\gamma_{\text{cusp}}\log\left(\frac{p_T^2R^2}{\mu^2}\right) + \gamma^{{\rm J}_k}, ~~~~~~~~{\rm with~}
  \gamma^{{\rm J}_k}_0 = -2\gamma^k_0.
\end{align}
At one-loop level, it is the same as the one for the unmeasured jet function defined in \cite{Ellis:2010rwa}. In our framework, $Z^{{\rm J}}$ is given by  $ \langle \bm{Z}_{12}^{{\rm J}} \hat \otimes \bm{1} \rangle$ at this order, where $\bm{Z}_{12}^{{\rm J}}$ removes the divergence in the jet function $\bm{\mathcal{ J}}_2$.  However, beyond one-loop level a simple correspondence between $Z^{{\rm J}} $ and the renormalization constants of the unmeasured jet function does not exist \cite{Becher:2016mmh}. 

Finally, we will discuss the NGL resummation. At the NLL level, the non-global evolution matrix from the coft scale to the jet scale reduces to
\begin{align}
    U_{\rm NG}(\mu_t,\mu_j) \xrightarrow[]{\text{NLL}} \sum_{m\geq 1}^\infty  \big\langle  \bm{U}_{1m}(\{\underline{n}\},\mu_t,\mu_j)\,\hat{\otimes}\, \,
    \bm{1}\big\rangle 
\end{align}
where we truncate the first sum in \eqref{eq:UNG} at the tree-level jet function $\bm{\mathcal{ J}}_1=4\pi\delta^{(d-2)}(\vec{n}_{J_1\perp})\bm{1}$, and we only include the tree-level coft function $\bm{\mathcal{ U}}_m=\bm{1}$. The non-global anomalous dimension defined in \eqref{eq:JUrg} has the following form \cite{Becher:2016mmh}
\begin{align}
\hat{\bm{\Gamma}}_{lm} (\{\underline{n}\}) = \frac{\alpha_s}{4\pi} \left( \begin{array} { c c c c c } { \overline {\boldsymbol  V} _ { 1 } } & { \boldsymbol { R } _ { 1 } } & { 0 } & { 0 } & { \ldots } \\ { 0 } & { \overline{\boldsymbol  V } _ { 2 } } & { \boldsymbol { R } _ { 2 } } & { 0 } & { \ldots } \\ { 0 } & { 0 } & { \overline{\boldsymbol  V } _ { 3 } } & { \boldsymbol { R } _ { 3 } } & { \ldots } \\ { 0 } & { 0 } & { 0 } & { \overline {\boldsymbol  V } _ { 4 } }& {\ldots } \\ { \vdots } & { \vdots } & { \vdots } & { \vdots } & { \ddots } \end{array} \right) + \mathcal{O}(\alpha_s^2)
\end{align}
where we only need the one-loop results for the NLL resummation. The matrix elements are given by
\begin{align}
 \label{Vanom}\overline {\boldsymbol  V} _ { m } = & \, 2 \sum _ {  i, j  } \left( \boldsymbol { T } _ { i , L } \cdot \boldsymbol { T } _ { j , L } + \boldsymbol { T } _ { i , R } \cdot \boldsymbol { T } _ { j , R } \right) \int \frac { d \Omega \left( n _ { k } \right) } { 4 \pi } W _ { i j } ^ { k }   \notag\\ 
& -  2 \left( \boldsymbol { T } _ { 0 , L } \cdot \boldsymbol { T } _ { 1 , L } + \boldsymbol { T } _ { 0 , R } \cdot \boldsymbol { T } _ { 1 , R } \right) \int \frac { d \Omega \left( n _ { k } \right) } { 4 \pi } W _ { 01 } ^ { k }   \Theta _ { \mathrm { out } } ( n_k ),  \\
  \boldsymbol { R } _ { m } = & - 4 \sum _ {  i, j  } \boldsymbol { T } _ { i , L } \cdot \boldsymbol { T } _ { j , R } W _ { i j } ^ { m + 1 } \Theta _ { \mathrm { in } }  \left( n _ { m + 1 } \right) ,
\end{align}
where $\boldsymbol { T } _ { i ,L }$ are the color generators acting on the $i$-th particle in the amplitude and $\boldsymbol { T } _ { i ,R }$ are the ones acting on the conjugate amplitude. The angular dipole factor $W_{ij}^k$ is defined as
\begin{align}
W_{ij}^k = \frac{n_i \cdot n_j}{n_i \cdot n_k n_k \cdot n_j}. 
\end{align}
The second line in \eqref{Vanom} corresponds to the global anomalous dimension subtracted out from $\overline {\boldsymbol  V} _ { m }$, where we define $n_0=\bar n_J$.  By expanding $U^k_{\rm NG}(\mu_t, \mu_j)$ as a series of
\begin{align}
t \equiv  \int_{\mu_t}^{\mu_j} \frac{d\mu}{\mu}  \frac{\alpha_s(\mu)}{4\pi},
\end{align}
we have the evolution factor $U_{\rm NG}(\mu_t, \mu_j) = t \, \bm{\mathcal U}_1^{(1)} + t^2 \bm{\mathcal U}_1^{(2)} + \cdots $ with one- and two-loop coefficients as
\begin{align}\label{eq:ngl}
\bm{\mathcal U}_1^{(1)} = & 0, \\
\bm{\mathcal U}_2^{(1)} = & - 16 C_k C_A \int \frac{d\Omega(n_2)}{4\pi} \frac{d\Omega(n_3)}{4\pi} \Theta_{\rm in}(n_2) \Theta_{\rm out}(n_3) \left[ W_{01}^2(W_{12}^3 + W_{02}^3) -W_{01}^2W_{01}^3 \right]. \notag 
\end{align}
From this, one can show that the coefficient of the leading NGL at two loops is $-4 C_k C_A \pi^2/3$, which is the same as the results in \cite{Dasgupta:2001sh}.  As shown in \cite{Becher:2016mmh,Dasgupta:2012hg}, the coft function maps onto the hemisphere soft function under a Lorentz boost along the jet axis. Therefore, the evolution of the function $U_{\rm NG}$ should be the same as Dasgupta and Salam's parametrization in \cite{Dasgupta:2001sh}. Explicitly, in our numerical calculations we have
\begin{align}\label{eq:NGL}
U^k _ { \mathrm { NG } } \left( \mu _ { t } , \mu _ { j } \right) \approx \exp \left( - C _ { A } C _ { k } \frac { \pi ^ { 2 } } { 3 } u ^ { 2 } \frac { 1 + ( a u ) ^ { 2 } } { 1 + ( b u ) ^ { c } } \right).
\end{align}
Here $u = 2 t = \frac{1}{\beta_0} \log \frac{\alpha_s(\mu_t)}{\alpha_s(\mu_j)}$, and the constants are given as $a = 0.85 C_A$, $b=0.86 C_A$ and $c=1.33$.

\subsection{NLL resummed expression}
After plugging in with the above expressions, the all-order resummed expression \eqref{eq:dsigmaResumFinal} could be reduced to
\begin{align}\label{eq:dsigmaNLL}
&\frac{d\sigma^{\rm NLL}}{d^2q_{T} d^2 p_{T} d\eta_J dy_V}=\sum\limits_{ijk}\int \frac{d^2 x_T}{(2\pi)^2} e^{i\vec{q}_{T}\cdot \vec{x}_T} B_{i/N_1}(\xi_1,x_T,\mu)B_{j/N_2}(\xi_2,x_T,\mu)\left( \frac{x_T^2 \, \hat s}{b_0^2} \right)^{-(C_i + C_j){F_\perp(\mu)}} \notag\\
&~~\times \exp\left[\int_{\mu_h}^{\mu} \frac{d \bar \mu}{ \bar \mu} \Gamma^{{\rm H}_{ij\to V k}}( \bar \mu) +\int_{\mu_b}^{\mu} \frac{d \bar \mu}{\bar \mu} \Gamma^{{\rm W}_{ij\to V k}}( \bar \mu) + \int_{\mu_j}^{\mu} \frac{d\bar\mu}{ \bar\mu} \Gamma^{{\rm J}_{k}}( \bar\mu) + \int_{\mu_t}^{\mu} \frac{d\bar \mu}{\bar \mu} \Gamma^{{\rm U}_{k}}( \bar \mu) \right] \notag \\
&~~\times\mathcal{H}_{ij\to Vk}(\hat s,\hat t,m_V,\mu) U_{\rm NG}^k(\mu_t,\mu_j).
\end{align}
Let us compare our NLL resummed expression with those in \cite{Chen:2018fqu, Sun:2018icb, Buffing:2018ggv}. The large logarithms $\log(Q/q_T)$ were resummed in $\gamma$+jet \cite{Chen:2018fqu} and $Z$+jet \cite{Sun:2018icb} events at NLL level using the CSS formalism \cite{Collins:1984kg}. In these references, the calculations were carried out in the small $R$ limit, but the terms with $\log R$ in the coefficients are not completely resummed. NGLs were also neglected. In the effective field theory language, this simply means that one does not distinguish the $n_J$-collinear mode from the hard mode, and also not distinguishing the coft mode from the soft mode. By taking $\mu_j=\mu_h$, $\mu_t=\mu_b$ and switching off the NGL resummation, our resummed expression reduces to those used in \cite{Chen:2018fqu,Sun:2018icb}. This can be shown more explicitly if one takes $\mu=\mu_b$ and 
\begin{align}\label{eq:BBf}
B_{i/N_1}(\xi_1,x_T,\mu_b){B}_{j/N_2}(\xi_2,x_T,\mu_b)= f_{i/N_1}(\xi_1,\mu_b)f_{j/N_2}(\xi_2,\mu_b).
\end{align}
On the other hand, in \cite{ Buffing:2018ggv} the authors performed a resummation of $\log R$ and $\log (Q/q_T)$ without resumming non-global logarithms. If we take $U_{\rm NG}^k=1$, our resummed expression formally reduces to their results.

\section{Analysis of Leading Logarithms}\label{sec:LL}
 
In this section we analyze the LL resummation. We shall study two cases of $p_T \gtrsim m_V$ and $p_T \lesssim m_V$, respectively. The first case is relevant in the studies of $\gamma$+jet production since photon is massless, or massive boson +jet production at high $p_T$, while the second case is relevant in massive boson+jet production at low $p_T$. Since leading logarithms are insensitive to scale choice, we choose $\mu_b=b_0/x_T, \mu_t=R\, b_0/x_T$ and $\mu_j=R\, p_T$ in the following discussions.
  
 \subsection{Leading Logarithms for $p_T \gtrsim m_V$}\label{sec:LLhighpT}
 
In this case all the collinear particles typically carry an energy of order $ p_T$. Therefore one can simply make the following replacement
 \begin{align}
 \mu_h^2\to p_T^2,~~~~ \hat{s}\to p_T^2,~~~~ -\hat{u}\to p_T^2,~~~~-\hat{t}\to p_T^2. 
 \end{align}
Then at LL level, (\ref{eq:dsigmaResumFinal}) reduces to the following form
\begin{align}\label{eq:LL}
\frac{dn(q_T)}{dq_T}&\equiv 2\pi q_T\int \frac{d^2x_T}{(2\pi)^2}e^{i \vec{x}_T\cdot \vec{q}_T} e^{-\frac{\alpha_s}{\pi}\left[(C_i+C_j)\log^2\left(\frac{p_T x_T}{b_0}\right) +C_k\log\left(\frac{1}{R^2}\right)\log\left(\frac{p_T x_T}{b_0}\right) \right]}\notag\\
&= \frac{2\alpha_s}{\pi q_T}\left[(C_i+C_j)\log\left(\frac{p_T}{q_T}\right) +C_k\log\left(\frac{1}{R}\right) \right] e^{-\frac{\alpha_s}{\pi}\left[(C_i+C_j)\log^2\left(\frac{p_T }{q_T}\right) +C_k\log\left(\frac{1}{R^2}\right)\log\left(\frac{p_T }{q_T}\right) \right]},
\end{align}
where $dn(q_T)/dq_T$ is the differential probability of the boson+jet transverse momentum $q_T$. We have only kept the LL terms in performing the Fourier transformation by using the relation
\begin{align}
\text{Fourier transform of $\log^n\left(\frac{p_T x_T}{b_0}\right)$}\to-\frac{1}{2\pi q_T^2} n \log^{n-1}(p_T/q_T).
\end{align}
We find that the resummation formula used in \cite{Chen:2018fqu, Sun:2018icb} give the same result at LL level. 

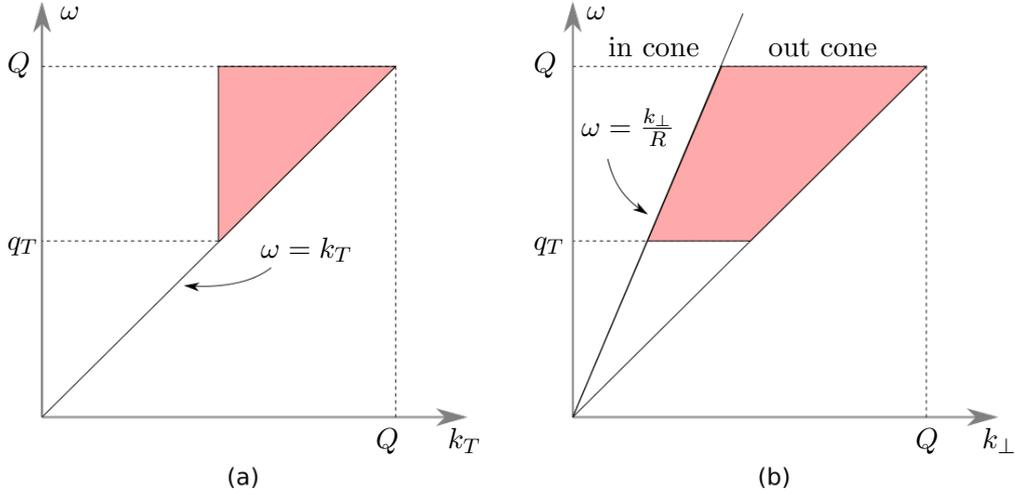
\begin{figure}[t!]

\begin{center}
\begin{overpic}[width=0.85\textwidth]{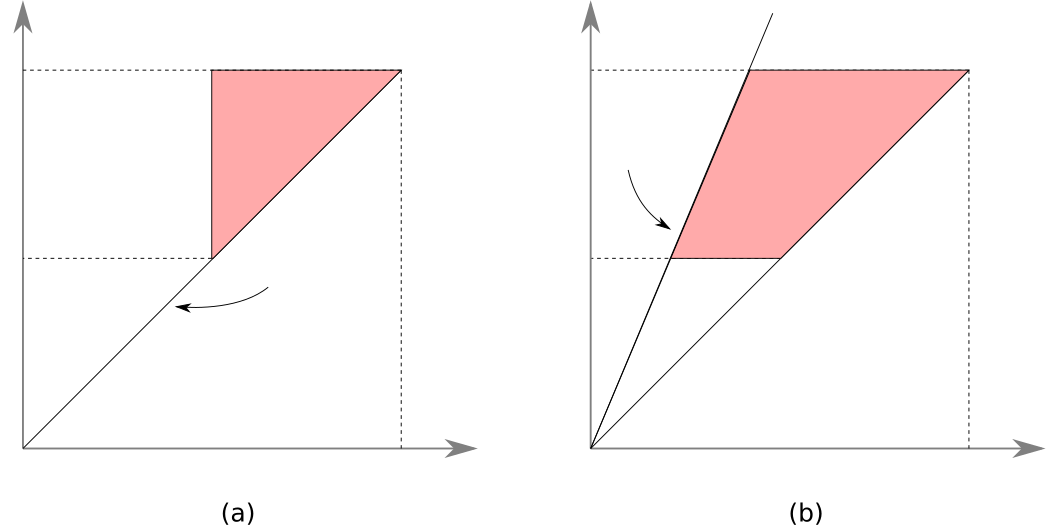}
\put(15,177){$\omega$}
\put(-5,157){$Q$}
\put(-5,90){$q_T$}
\put(90,87){$\omega=k_T$}
\put(133,15){$Q$}
\put(160,15){$k_T$}

\put(212,177){$\omega$}
\put(192,157){$Q$}
\put(192,90){$q_T$}
\put(210,133){$\omega=\frac{k_\perp}{R}$}
\put(220,163){in cone}
\put(280,163){out cone}
\put(335,15){$Q$}
\put(360,15){$k_\perp$}
\end{overpic}
\hspace{-3cm}
\end{center}
\caption{The double logarithmic phase-space for soft radiation along the two beam directions (a) and the jet direction (b).}\label{fig:phasespace}
 \end{figure}

The double logarithms in  (\ref{eq:LL}) arise from soft radiation along the three collinear directions. Using a physical gauge, such as the light-cone gauge, the soft gluon spectrum is given by
\begin{align}
\frac{dI}{d\omega \, dk_{n_a\perp}}=\frac{2\alpha_s}{\pi} C_a \frac{1}{\omega}\frac{1}{k_{n_a\perp}}.
\end{align}
Let us first calculate the beam contributions to the integral distribution $n(q_T)$. The phase space constraint for the soft gluon is as follows, 
\begin{align}
\begin{array}{ll}
k_T\lesssim q_T,~k_T\lesssim \omega\lesssim p_T~~~~~~&\text{for real emissions},\\
k_T\lesssim \omega\lesssim p_T~~~~~~&\text{for virtual contributions}.
\end{array}
\end{align}
The real and virtual cancellation yields the phase space shown as the shaded region in figure \ref{fig:phasespace} (a) with $Q=p_T$, and this gives
\begin{align}
I_{a}=-\frac{2\alpha_s}{\pi} C_{a} \int^{p_T}_{q_T}\frac{dk_T}{k_{T}}\int _{k_T}^{p_T}\frac{d\omega}{\omega}=-\frac{\alpha_s}{\pi} C_a\log^2\left(\frac{p_T}{q_T}\right)\qquad \text{for }a=1,2.
\end{align}
On the other hand, soft radiation along the jet direction results in the $\log R$-dependent terms. For simplicity, we assume that the jet is central. If a gluon is emitted inside the jet, there is no additional constraint on its phase space since it does not change the value of $q_T$. If the gluon is emitted outside the jet, its energy has to satisfy $\omega \lesssim q_T$. Combined with the virtual contribution, one can see that the phase space of the gluon is given by figure \ref{fig:phasespace} (b), which gives
\begin{figure}[t]
 \centering
\includegraphics[width=0.48\textwidth]{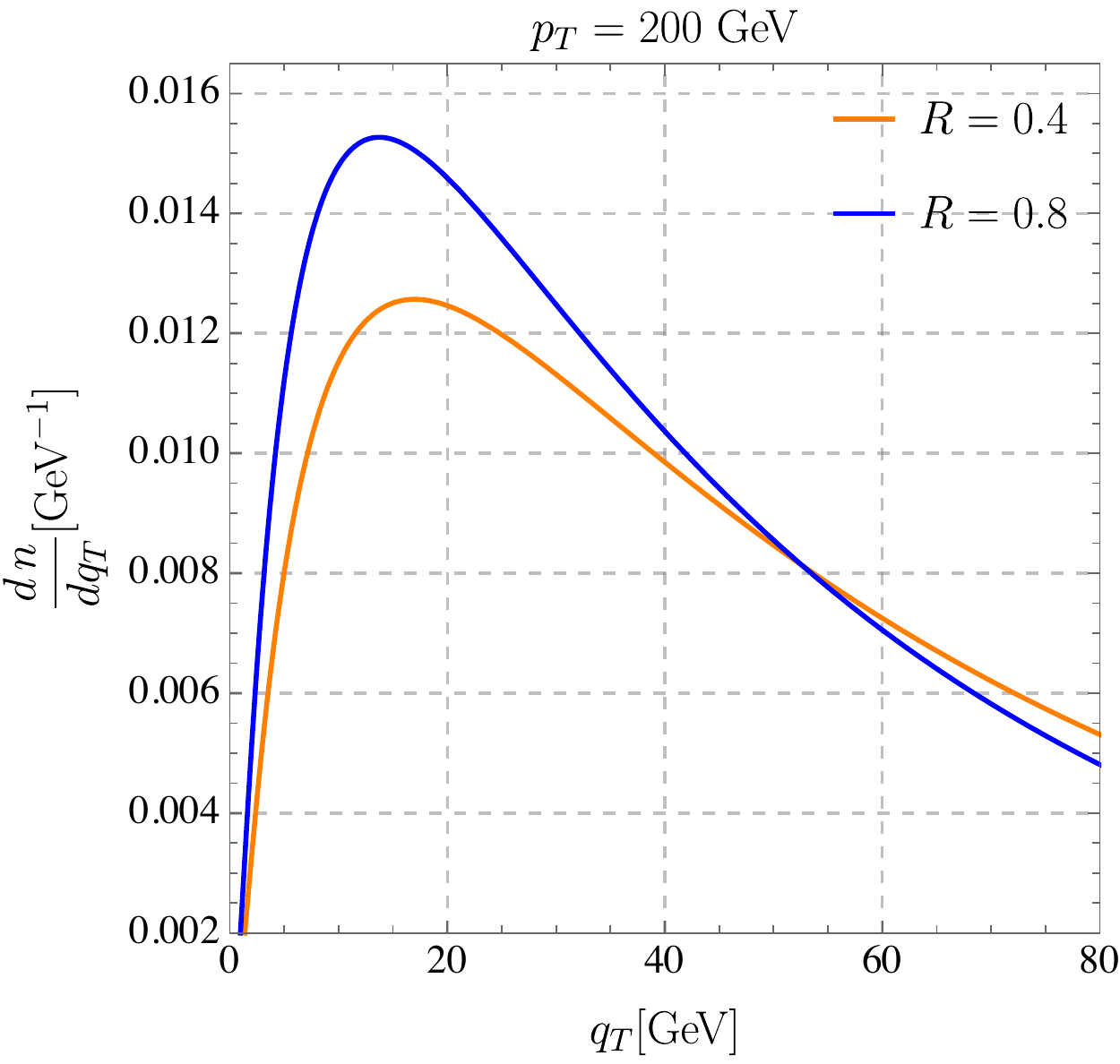} ~~
\includegraphics[width=0.47\textwidth]{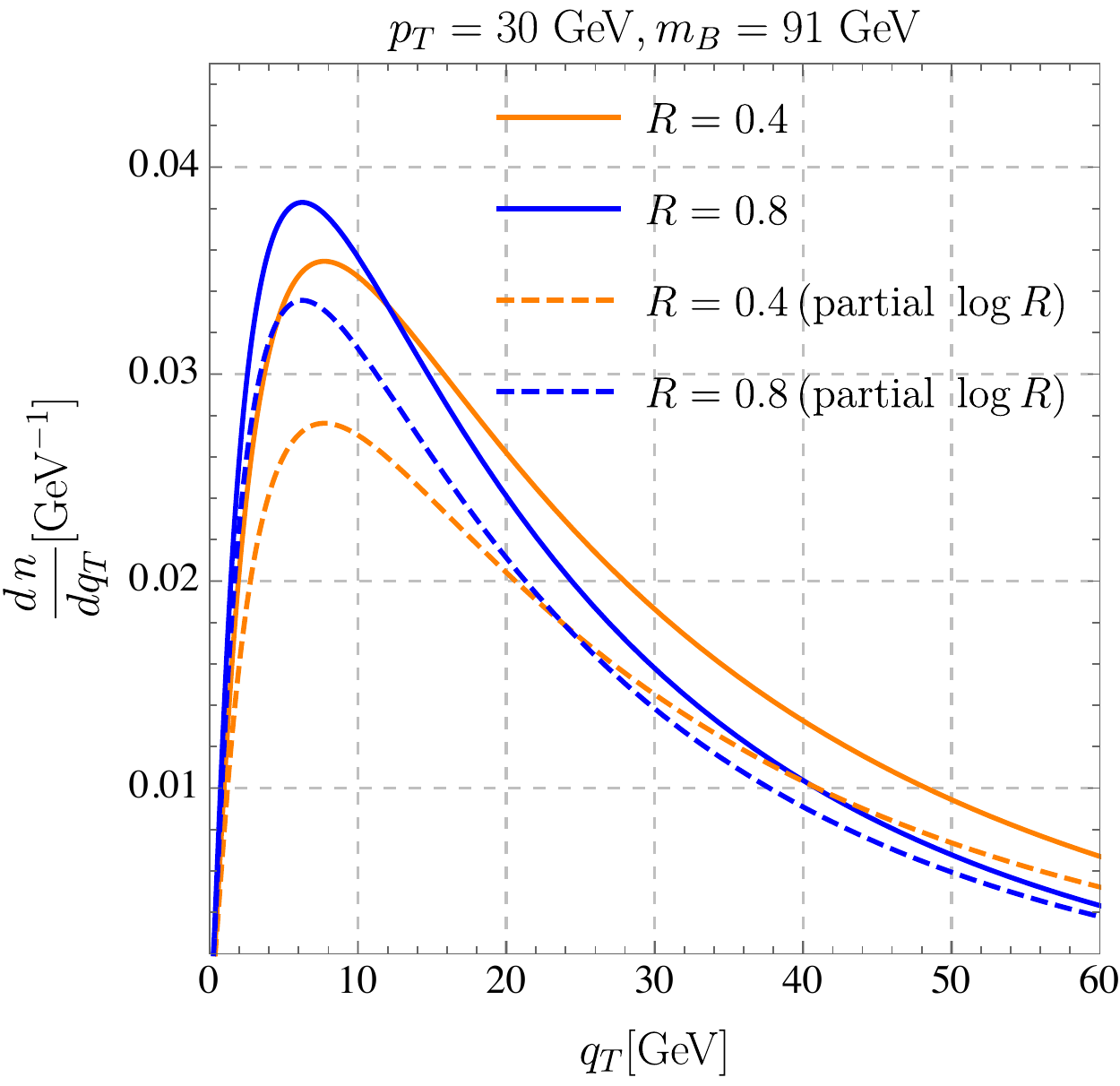}
\caption{Illustration of the $\log R$ dependence in $dn(q_T)/dq_T$ for the $qg$ channel at LL. Here we take $\alpha_s(q_T)\approx 0.18$ with $q_T\approx 10~$GeV around the peak region.
}\label{fig:LL}
\end{figure}
\begin{align}
I_{k}=-\frac{2\alpha_s}{\pi} C_k \int _{q_T}^{p_T}\frac{d\omega}{\omega}\int^{\omega}_{\omega R}\frac{dk_T}{k_{T}}=-\frac{\alpha_s}{\pi} C_k\log\left(\frac{1}{R^2}\right) \log\left(\frac{p_T}{q_T}\right).
\end{align}
By including uncorrelated multiple soft gluon radiation, one obtains the Sudakov factor of the form
\begin{align}\label{eq:nqT}
n(q_T)=e^{-\frac{\alpha_s}{\pi}\left[(C_i+C_j)\log^2\left(\frac{p_T }{q_T}\right) +C_k\log\left(\frac{1}{R^2}\right)\log\left(\frac{p_T }{q_T}\right) \right]}.
\end{align}
It is easy to see that differentiating $n(q_T)$ with respect to $q_T$ gives  (\ref{eq:LL}), which brings down from the exponent a factor given by single-gluon emission along the three collinear directions.  Note that $n(q_T=p_T)=1$ which recovers the whole probability. Also,  $dn(q_T)/dq_T$ peaks at
\begin{align}
\frac{q_T}{p_T}=e^{-\frac{1+\sqrt{1+2 \bar\alpha}}{\bar \alpha}} R^{-\frac{C_k}{C_i+C_j}}
\end{align}
with $\bar \alpha=4(C_i+C_j)\frac{\alpha_s}{\pi}$.  The peak location moves to a larger value of $q_T$ for smaller $R$ because the probability becomes larger for a gluon to be emitted outside the jet. Also, the height of the peak
\begin{align}\label{eq:LLpeak}
 \frac{1}{2} \left(\sqrt{2 \bar \alpha+1}+1\right) e^{\frac{2 (1+\frac{\alpha_s}{\pi}C_k \log R^2)^2-\bar \alpha+\sqrt{2\bar \alpha+1}-1}{2\bar \alpha}},
\end{align}
becomes lower. All the above-mentioned features are illustrated in the left plot of figure \ref{fig:LL}, which shows $dn(q_T)/dq_T$ with $R=0.4$ and 0.8, and we set  $\alpha_s=0.18$\footnote{As we will show in the next section, the peak of the differential $q_T$ distribution locates at $q_T\approx 10~$GeV, and $\alpha_s(10~\text{GeV})\approx 0.18$.}.

\subsection{Leading Logarithms for $p_T\lesssim m_V$}\label{sec:LLlowpT}
 
In this case the collinear radiation along the two beam directions typically has an energy $\sim m_V$, while the energy of the collinear particles inside the jet is of order $\sim p_T$. As a result, the phase space for soft radiation collinear to the jet direction is unmodified as in the previous case in figure \ref{fig:phasespace} (b). On the other hand, $m_V$, instead of $p_T$, sets the phase space of soft radiation along the two beam directions, which is given by figure \ref{fig:phasespace} (a) with $Q=m_V$. Based on this physical argument, one expects the following logarithms to show up in the calculation: $-\frac{\alpha_s}{\pi}(C_i+C_j)\log^2(m_V/q_T)$ and $\frac{2\alpha_s}{\pi}C_k\log(p_T/q_T)\log R$.
 
At LL accuracy,  using  (\ref{eq:dsigmaResumFinal}) one can simply set
\begin{align}
 \mu_h^2\to m_V^2,~~~~ \hat{s}\to m_V^2,~~~~ -\hat{u}\to p_T m_V,~~~~-\hat{t}\to p_T m_V.
\end{align}
Plugging them into  (\ref{eq:dsigmaResumFinal}), we get 
\begin{align}\label{eq:LLlowpT}
\frac{dn(q_T)}{dq_T}\equiv 2\pi q_T\int \frac{d^2x_T}{(2\pi)^2}&e^{i \vec{x}_T\cdot \vec{q}_T} e^{-\frac{\alpha_s}{\pi}\left[(C_i+C_j)\log^2\left(\frac{m_V x_T}{b_0}\right) -C_k\log^2\left(\frac{m_V}{p_T}\right)+C_k\log\left(\frac{1}{R^2}\right)\log\left(\frac{p_T x_T}{b_0}\right) \right]}\notag\\
&= \frac{2\alpha_s}{\pi q_T} e^{\frac{\alpha_s}{\pi}C_k \log^2\frac{m_V}{p_T}} \left[(C_i+C_j)\log\left(\frac{m_V}{q_T}\right) +C_k\log\left(\frac{1}{R}\right) \right]\notag\\
&\times e^{-\frac{\alpha_s}{\pi}\left[(C_i+C_j)\log^2\left(\frac{m_V }{q_T}\right) +C_k\log\left(\frac{1}{R^2}\right)\log\left(\frac{p_T }{q_T}\right) \right]}.
\end{align}
Note the additional, $q_T$-independent logarithms $\log^2(m_V/p_T)$ appearing in the resummed result, which gives an overall normalization constant. In contrast, the LL result in \cite{Sun:2018icb} is given by (\ref{eq:LL}) with $p_T$ replaced by $m_V$, which is different from our result (\ref{eq:LLlowpT}). Such differences can be seen in the right plot of figure \ref{fig:LL}, where we use the legend of ``partial $\log R$" to distinguish these two cases. Note that in the $p_T\ll m_V$ limit, the large logarithms of $\log (m_V/p_T)$ need to be properly resummed which requires a factorization of the hard sector at the two scales $m_V$ and $p_T$. We leave the study of constructing such an effective theory for future work.

\section{NLL Resummation and Phenomenology}

In this section, we study the $Z$+jet production in proton-proton collisions at $\sqrt{s}=13~$TeV in the high $p_T^J$ case ($p^J_T>200~$GeV) and the low $p_T^J$ case ($p^J_T>30~$GeV). We impose the constraint $|\eta_J|<2.4$ on the jet pseudo-rapidity and allow all values of boson rapidity. We then compare our theoretical predictions at NLL accuracy with \textsc{Pythia} simulations (version 8.2) \cite{Sjostrand:2014zea} and the CMS data \cite{Sirunyan:2018cpw,Chatrchyan:2013tna}. 

\subsection{Characteristic Scales and Numerical Evaluations}

We choose the following characteristic scales,
\begin{align}\label{eq:scaleChoice}
\mu_h = Q\equiv \sqrt{p_{T}^2+m_Z^2},~~~\mu_j =R\, p_{T},~~~\mu_b = \frac{b_0}{x_T},~~~\mu_t=R\frac{b_0}{x_T}.
\end{align}
Note that both $\mu_b$ and $\mu_t$ depend on $x_T$, and one needs to include nonperturbative contributions when these scales approach $\Lambda_{\rm QCD}$. 
We focus on the effects of resummation in perturbative QCD, and we simply impose an upper limit of $x_T < x^{\rm max}_{T}=1.5$ $\rm GeV^{-1}$ in the $x_T$-integral \cite{Neill:2015roa}. By varying $x^{\rm max}_{T}$ from $1$ $\rm GeV^{-1}$ to $3$ $\rm GeV^{-1}$, we find that the dependence of the $q_T$ distribution on $x^{\rm max}_{T}$ is negligible compared to the uncertainties from scale variation. On the other hand, in the large $q_T\gtrsim p_T$ region where 
$\mu_t>\mu_j$ and $\mu_b>{\rm min}(\mu_h,p_T)$, the effective theory is no longer valid and we set $\mu_t=\mu_j$ and $\mu_b={\rm min}(\mu_h,p_T)$. In this region we need to switch off resummation and match the resummed results with the fixed-order predictions. However, different matching schemes will introduce additional source of uncertainties. We focus on estimating the theoretical uncertainty from scale variation since it is the dominant uncertainty at NLL accuracy. We leave the detailed studies of fixed-order matching and next-to-next-to-leading logarithmic (NNLL) resummation for future work. 

\begin{figure}[t]
 \centering
\includegraphics[width=0.465\textwidth]{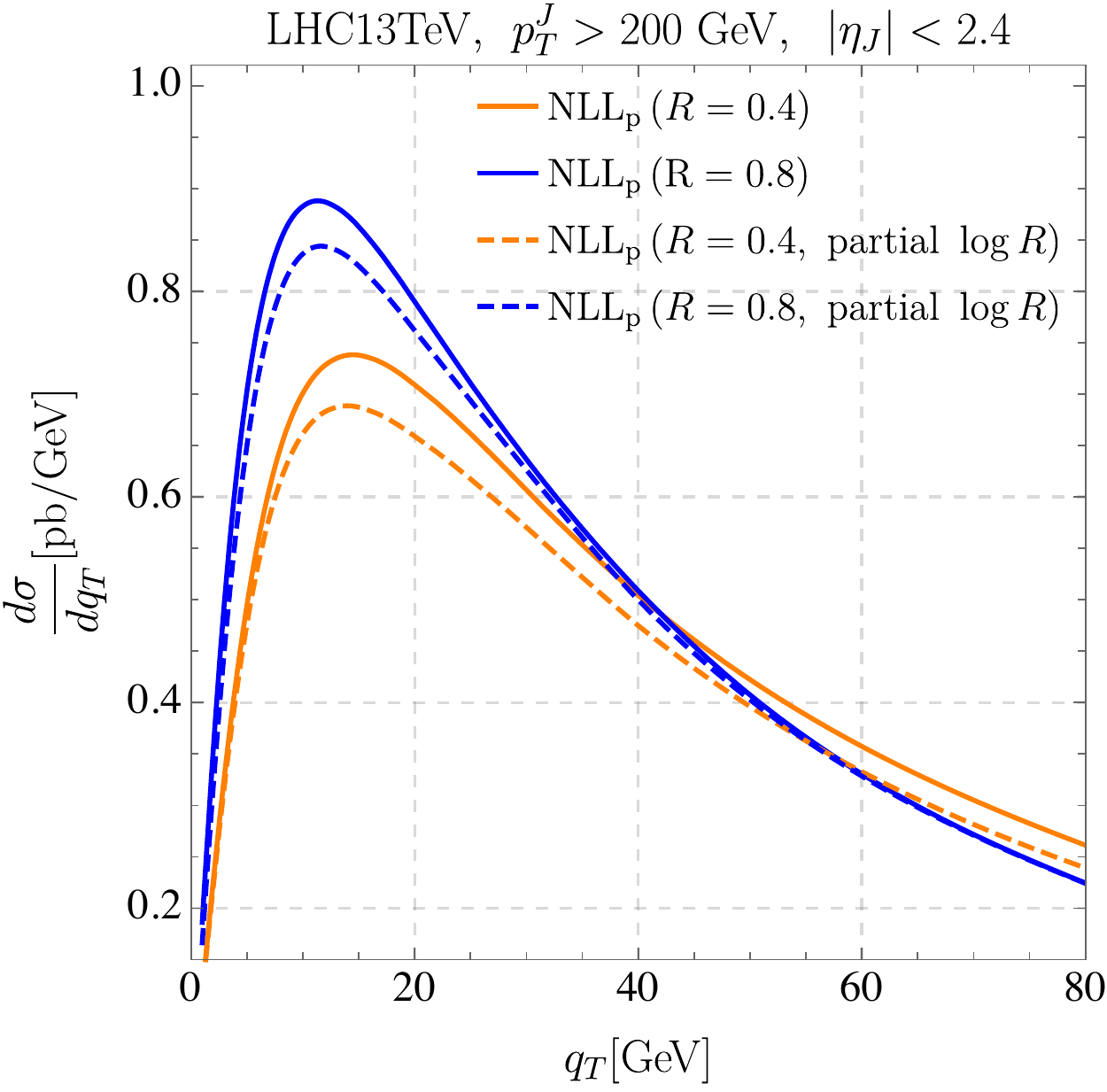} ~~
\includegraphics[width=0.47\textwidth]{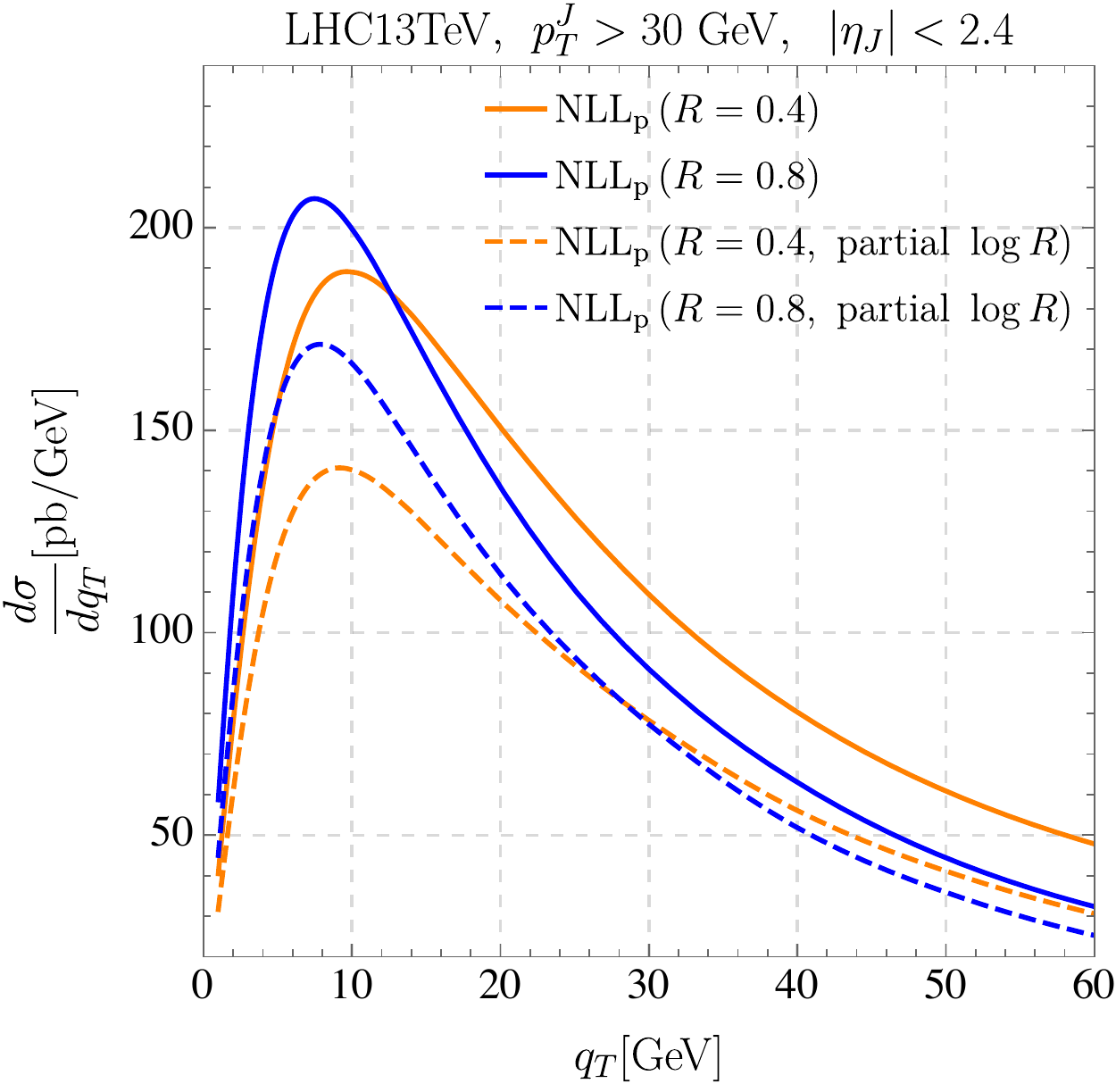}
\caption{Effects of $\log R$ resummation illustrated in the high $p_T^J$ (left plot) and low $p_T^J$ (right plot) cases. Here, NLL$_{\rm p}$ stands for the NLL resummed results excluding NGLs. The label ``partial log $R$" corresponds to the results from setting $\mu_j = \mu_h$ and $\mu_t = \mu_b$. }\label{Fig:qTNLLrResum}
 \end{figure}
The differential $q_T$ distribution $d\sigma/dq_T$ is calculated by numerically integrating over all the variables in (\ref{eq:dsigmaResumFinal}). For the NLL resummation, we need all the one-loop anomalous dimensions in section \ref{subsec:adoneloop} and the two-loop cusp anomalous dimension in appendix \ref{app:ad}. The exponential factors in (\ref{eq:dsigmaResumFinal}) are evaluated analytically according to (\ref{eq:exponential}) and (\ref{eq:SA}). We take the beam functions to be equal to the CT14 NLO PDF set \cite{Dulat:2015mca} at the scale $\mu_b$ according to (\ref{eq:BBf}). There is a constraint coming from requiring the $\phi_x$-integral to be convergent. Recall that both the soft and coft anomalous dimensions depend on $\cos\phi_x$. The $\phi_x$-dependent terms can be combined and factored out as,
\begin{align}
  |\cos\phi_x|^{\frac{4 C_k}{\beta_0}\log\frac{\alpha_s(\mu_b)}{\alpha_s(\mu_t)}}.
\end{align}
The $\phi_x$-integral is convergent only if 
\begin{align}\label{eq:phix}
-1<\frac{4 C_k}{\beta_0}\log\frac{\alpha_s(\mu_b)}{\alpha_s(\mu_t)}\approx -\frac{2\alpha_s(\mu_t)}{\pi}\log\frac{1}{R}.
\end{align}
One encounters such a divergence when the coft scale approaches to the non-perturbative region. It would be intriguing to see how one can introduce nonperturbative functions to tame such a divergence. We instead only integrate $x_T$ over the region given by \eqref{eq:phix}. 

\subsection{Effects of $\log R$ and Non-Global Logarithm Resummation} \label{sec:NLL}

We study the effects of $\log R$ and NGL resummation at NLL accuracy.

\subsubsection{$\log R$ resummation}

Here we switch off the contribution from NGLs by setting $U_{\rm NG}^k=1$ in (\ref{eq:dsigmaResumFinal}). We define the resummation accuracy without NGLs as NLL$_\text{p}$ where the subscript $_\text{p}$ means \textit{partial}. Furthermore, we compare the NLL$_\text{p}$ result with the one by setting $\mu_j=\mu_h$, $\mu_t=\mu_b$ which is denoted by ``partial $\log R$" resummation since part of the $\log R$ dependence is eliminated in the scale ratios. Note that, in the high $p_T$ case $\mu_j=\mu_h\sim p_T$ while in the low $p_T$ case $\mu_j=\mu_h\sim m_Z$, and that the characteristic scale $\mu_j=p_T R$.  Therefore in the low $p_T$ case the ``partial $\log R$" results differ from the NLL$_\text{p}$ result by the missing contributions of the form $\log (m_Z/(p_TR))$ as discussed in section \ref{sec:LLlowpT}. 

Figure \ref{Fig:qTNLLrResum} shows the effect of $\log R$ resummation in the high $p_T$ (left plot) and low $p_T$ (right plot) cases. The NLL$_\text{p}$ cross section is always larger than that with partial $\log R$ resummation. Note the significant effect on the overall cross section especially in the low $p_T$ case. As discussed in section \ref{sec:LL}, one can see that the overall factor
\begin{align}
e^{\frac{\alpha_{s}}{\pi} C_k \log \frac{m_Z}{p_T}\left(\log \frac{m_Z}{p_T}+2\log\frac{1}{R}\right)} 
\end{align}
accounts for the cross section difference, which clearly comes from the running of the jet function between $m_Z$ and $p_T R$.

\begin{figure}[t]
 \centering
\includegraphics[width=0.47\textwidth]{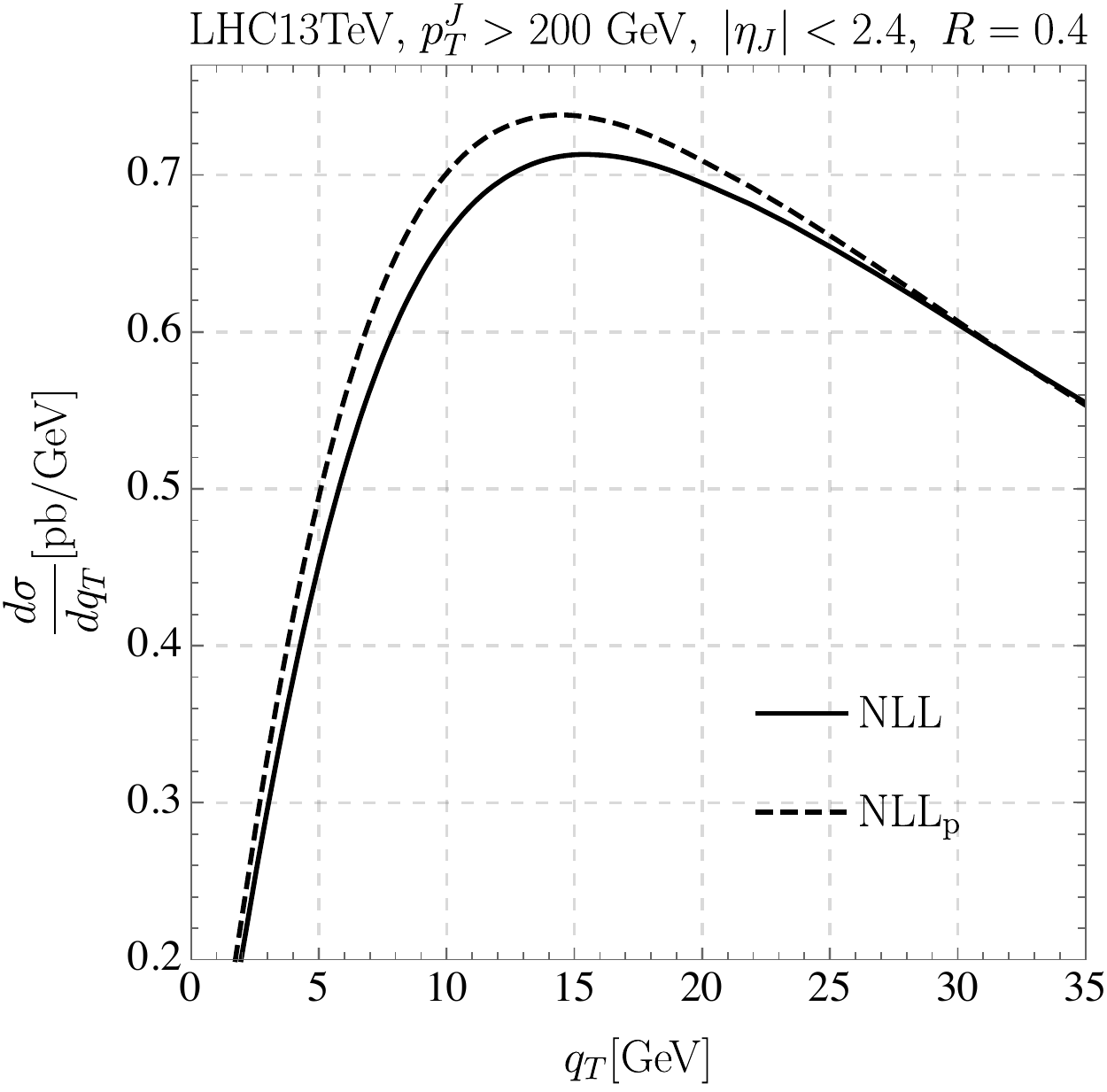} ~~
\includegraphics[width=0.47\textwidth]{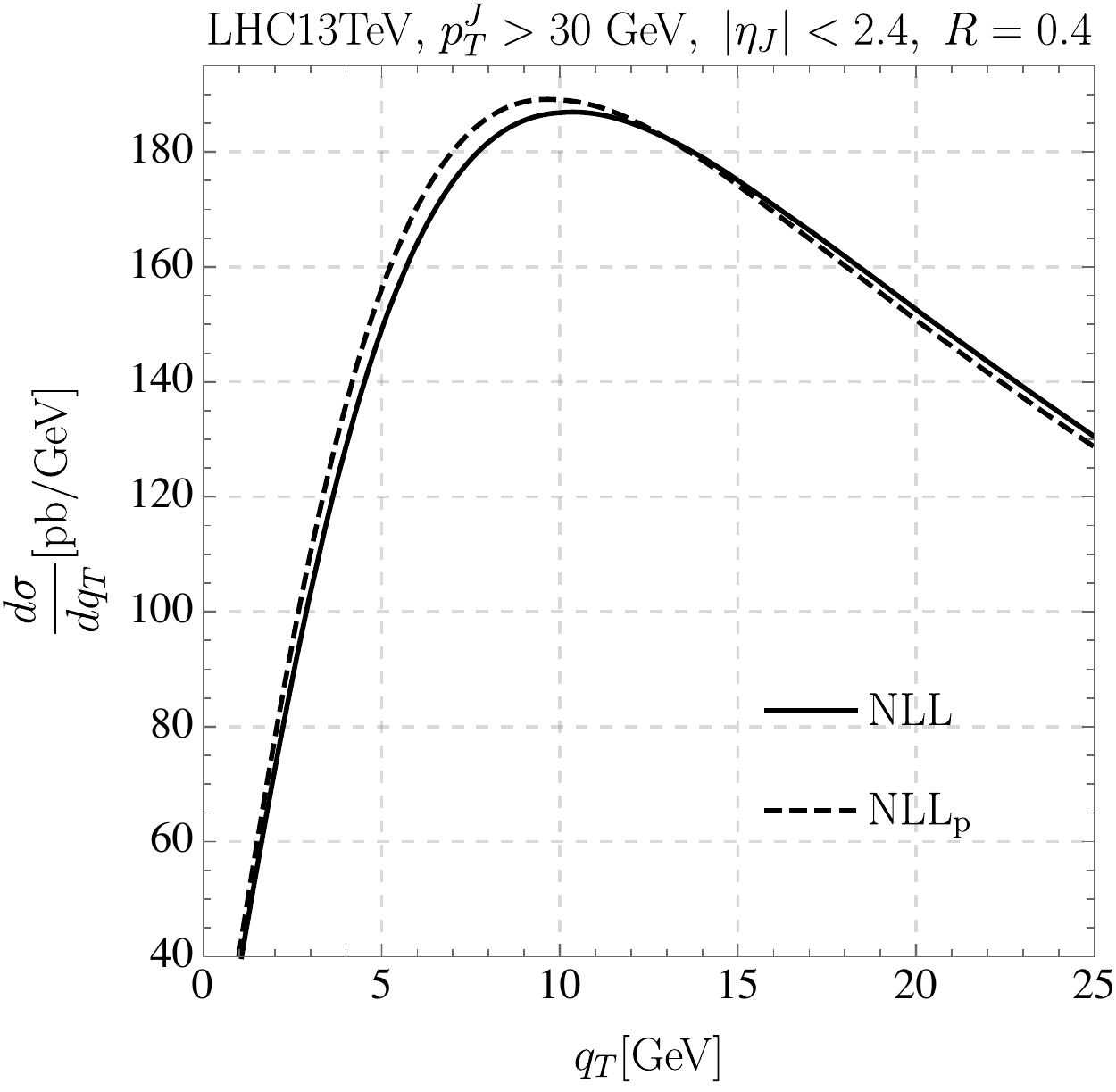}
\caption{The effect of NGL resummation illustrated in the high $p_T^J$ (left plot) and low $p_T^J$ (right plot) cases. Here, NLL stands for the full results calculated from (\ref{eq:dsigmaResumFinal}), and NLL$_\text{p}$ corresponds to the results without NGL resummation.}\label{Fig:qTNGL}
 \end{figure}

 \subsubsection{NGL resummation}

As discussed in section \ref{sec:resum}, NGLs arise from one coft gluon radiated outside the jet. The contribution at $O(\alpha_s^2)$ takes the form
\begin{align}\label{eq:NGLLO}
-\frac{\alpha_s^2}{12} C_A C_k \log^2\frac{\mu_j}{\mu_t}=-\frac{\alpha_s^2}{12} C_A C_k \log^2\frac{p_T}{q_T},
\end{align}
and the contribution increases as the ratio $p_T/q_T$ increases. Therefore NGLs are expected to play a more important role at high $p_T$. Figure \ref{Fig:qTNGL} shows the cross sections calculated from  (\ref{eq:dsigmaResumFinal}) with (denoted by NLL) or without (denoted by NLL$_\text{p}$) the NGL resummation. One can see that the NGL resummation lowers the peak of the cross section and pushes it to a larger value of $q_T$.

\subsection{Theoretical Predictions and Uncertainties}\label{sec:comparisonData}

We compare our theoretical predictions with \textsc{Pythia} simulations and experimental data at the LHC. We estimate the theoretical uncertainties by varying each characteristic scale in (\ref{eq:scaleChoice}) by a factor two and taking the envelope of all the results from scale variations. 

\begin{figure}[t]
 \centering
\includegraphics[width=1\textwidth]{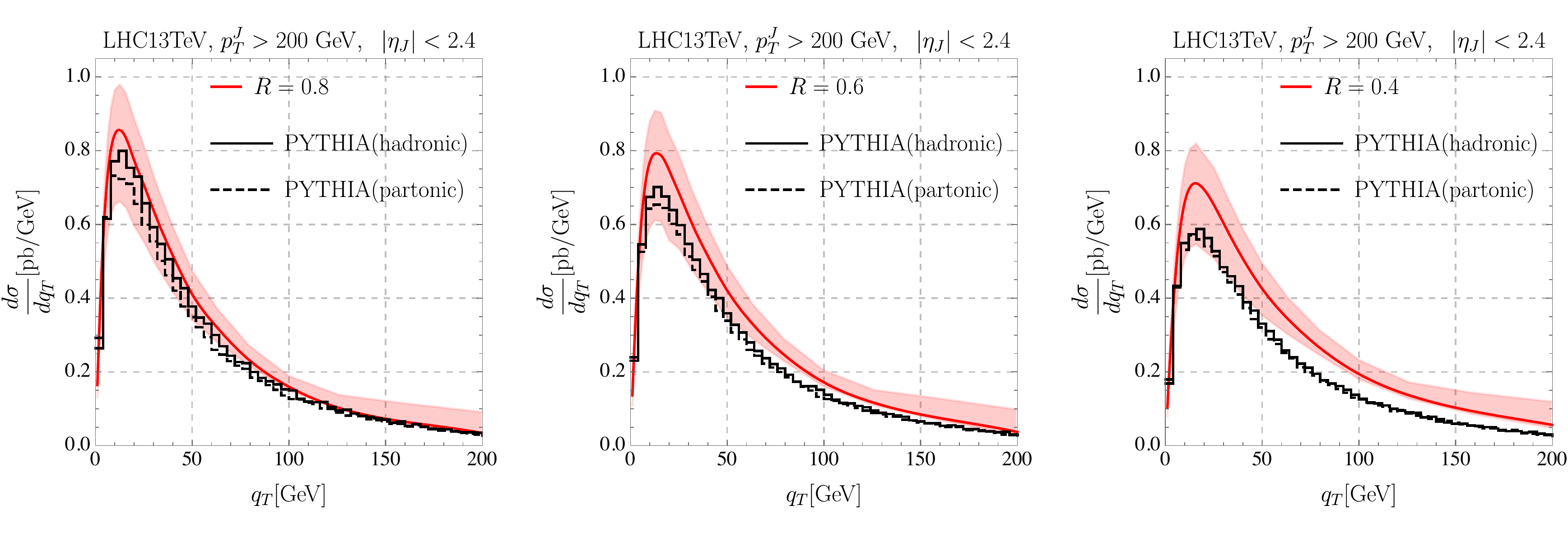}
\includegraphics[width=1\textwidth]{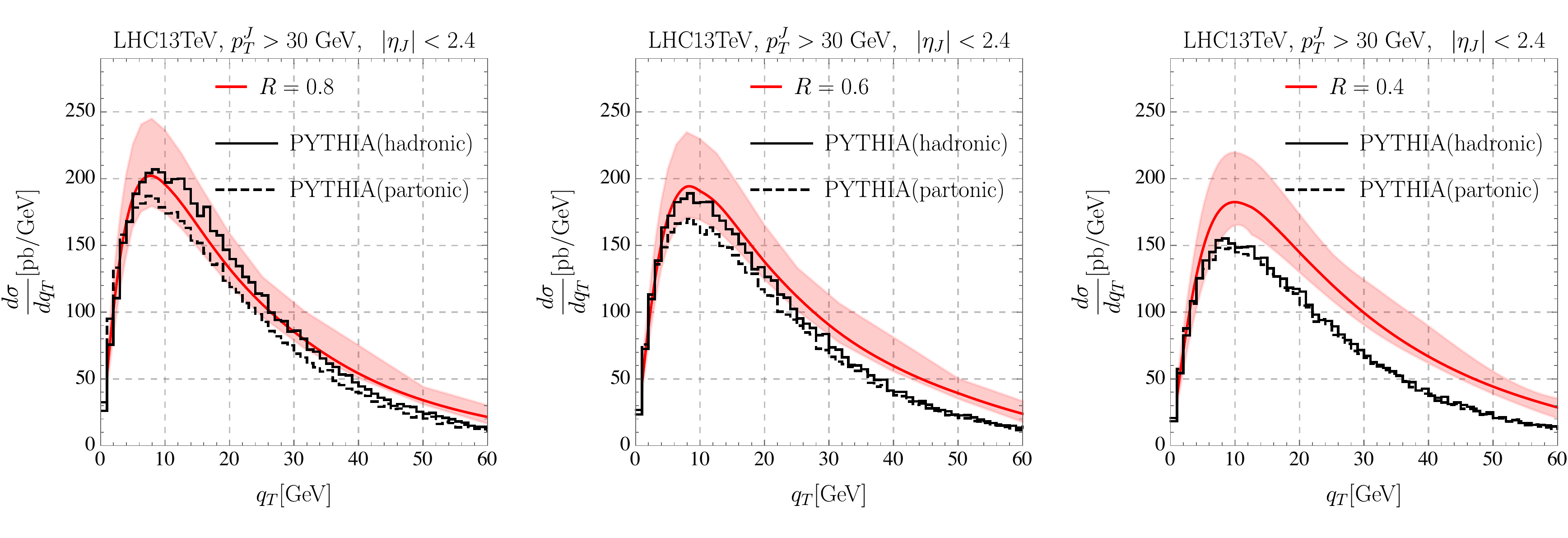}
\caption{Comparison between the NLL cross section calculations with \textsc{Pythia} simulations, in the high $p_T^J$ case (top row) and the low $p_T^J$ case (bottom row). In all the plots, the red curves are the theoretical predictions with the scale choice in (\ref{eq:scaleChoice}), and the error bands are shown as the shaded regions. The histograms are the \textsc{Pythia} results at parton (dashed lines) and hadron (solid lines) levels.}\label{Fig:dsigmadqT}
 \end{figure}

For both the $p_{T}>30$~GeV and $p_{T}>200$~GeV cases, we calculate $d\sigma(q_T)/dq_T$ with $R=0.4, 0.6$ and $0.8$. As shown in figure \ref{Fig:dsigmadqT}, our theoretical predictions agree reasonably well with the \textsc{Pythia} partonic results within the uncertainty band\footnote{We checked that the major difference between the partonic and hadronic results comes from multi-parton interaction contributions.}. 
However, some discrepancy in the overall cross section exists, especially for $R=0.4$ with a smaller jet radius. 

\begin{figure}[t]
 \centering
\includegraphics[height=0.45\textwidth]{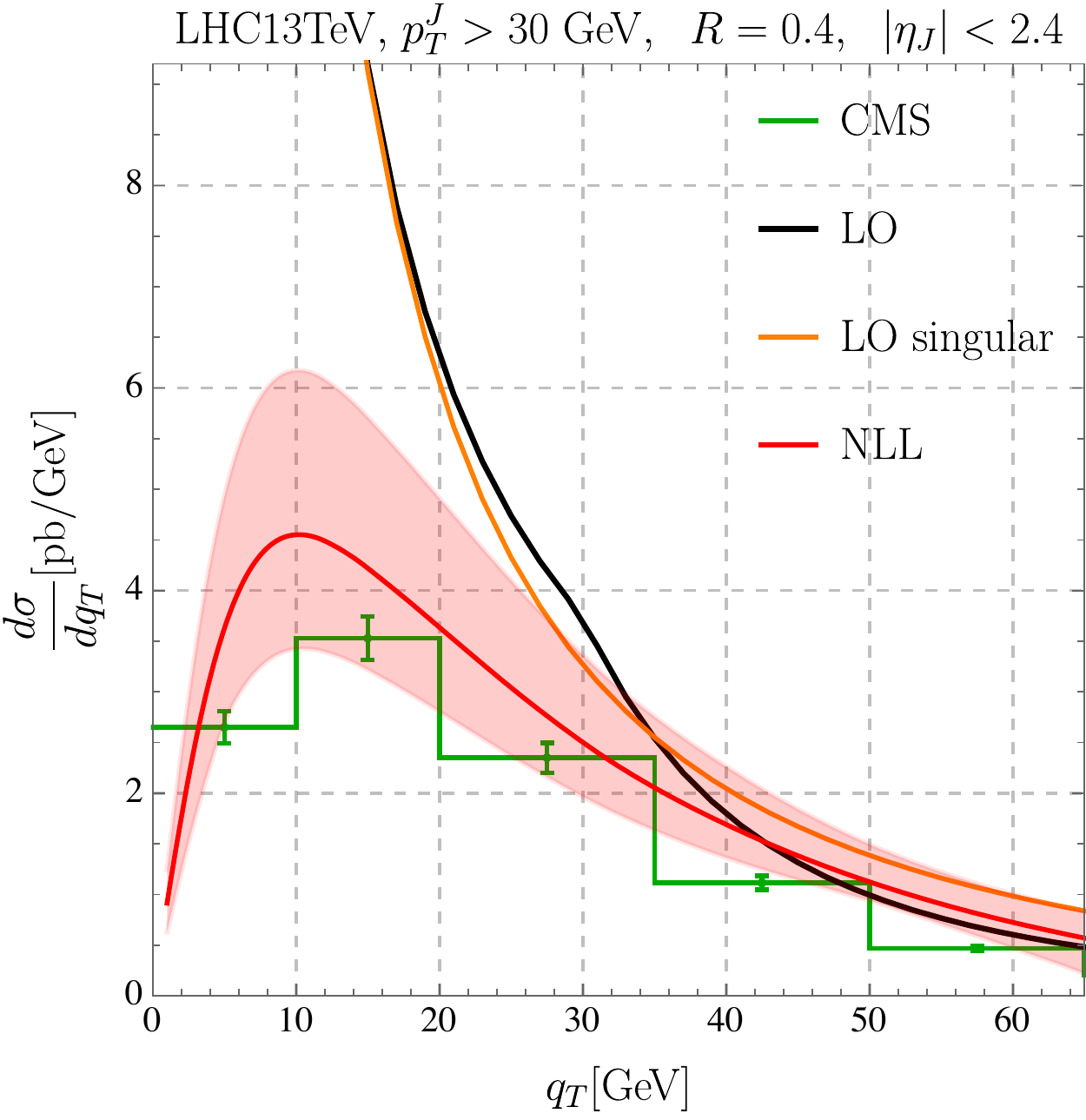} ~~~
\includegraphics[height=0.45\textwidth]{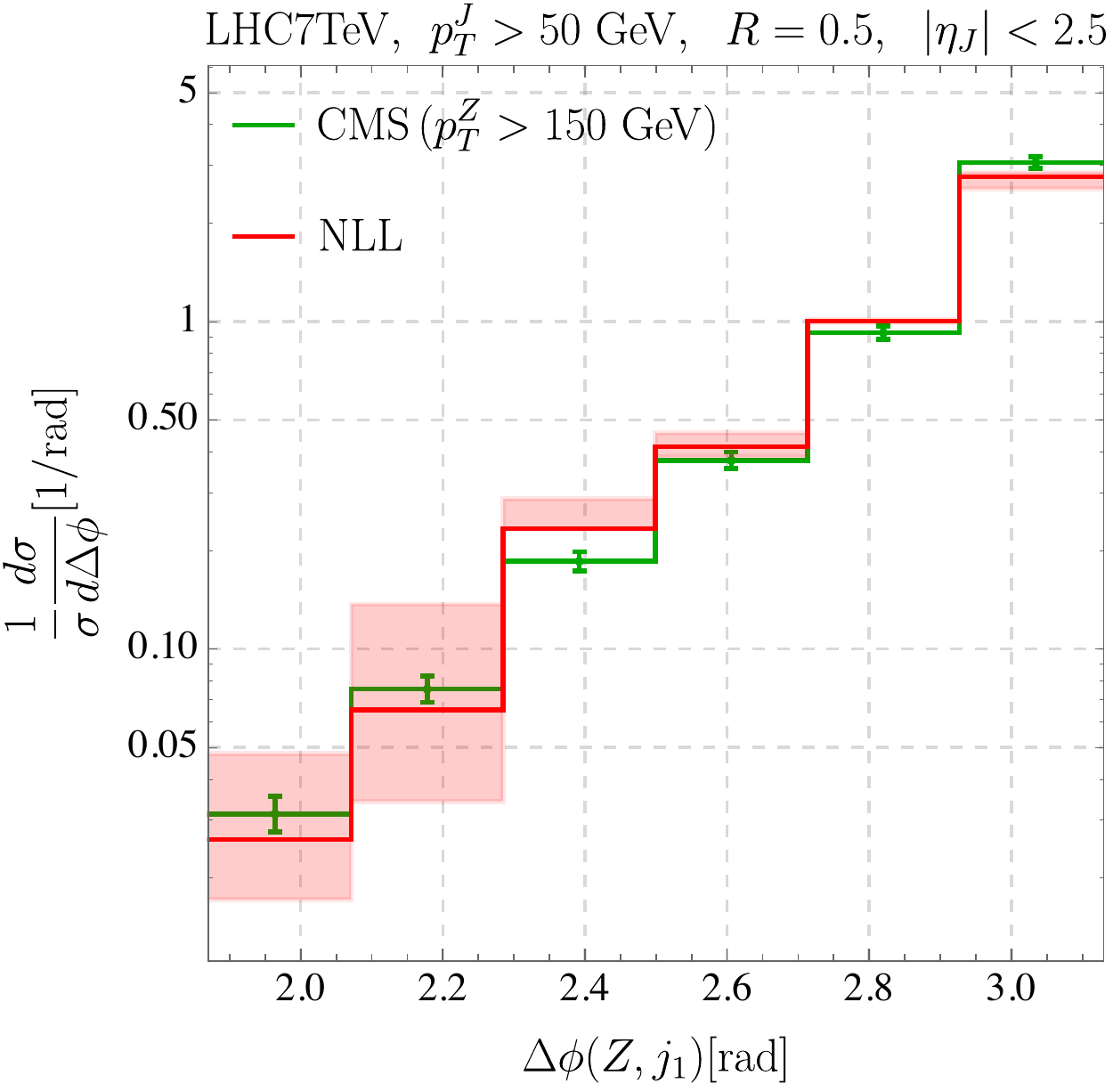}
\caption{Comparison between theoretical calculations with experimental data for the process: $pp\to Z^0/\gamma^*\to e^+ e^-$ and $\mu^+\mu^-$. The left plot shows the comparison between our NLL result of $d\sigma/dq_T$ with the measurement in  \cite{Sirunyan:2018cpw}, where the solid red curve is the result with the scale choice in (\ref{eq:scaleChoice}) and the shaded region indicates the error band from scale variation. The solid black and orange curves are respectively the LO result and the LO result including only logarithmic terms (LO singular). The right plot shows the comparison between our prediction of azimuthal angle decorrelation with the measurement in \cite{Chatrchyan:2013tna}, where $\Delta\phi(Z,j_1)$ is defined as the azimuthal angle between the $Z$ boson and the leading jet.}\label{Fig:qTCMS}
 \end{figure}

We then compare our theoretical calculation with experimental data. In order to impose the same cuts on kinematic variables as the experiments, we use the LO hard function including the leptonic decay of $Z/\gamma^*$. We first compare with the data at $\sqrt{s}=13~$TeV in \cite{Sirunyan:2018cpw}. We impose the same kinematic cuts as
\begin{align}
&p_{T}^J>30~{\rm GeV},~~|\eta_J|<2.4,~~R=0.4, \notag \\
&p_{T}^l>20~{\rm GeV},~~|\eta_l|<2.4,~~71~{\rm GeV}<m_{ll}<111~{\rm GeV}.
\end{align} 
The left plot of figure \ref{Fig:qTCMS} shows the comparison between our prediction for $d\sigma(q_T)/dq_T$ with the data\footnote{Note that in experiment $q_T$ is defined as the sum of the transverse momenta of the $Z$ boson and {\sl all} the jets with $p_T^J>30~$GeV and $|\eta_J|<2.4$ in the event \cite{Sirunyan:2018cpw}, while in our calculation we only include the leading jet in defining the $q_T$. From \textsc{Pythia} simulations, we find that using the leading jet to define $q_T$ brings down the first three bins of $d\sigma/dq_T$ (left plot of figure \ref{Fig:qTCMS}) by 6.2\%, 8.9\% and 5.7\%, respectively.}. 
Our result is consistent with the experimental data in the small $q_T$ region. We also show the result of the full LO distribution (black curve) calculated by MCFM program~\cite{Campbell:2002tg,Campbell:2003hd}, and the one including only the logarithmic terms at LO (orange curve) predicted using SCET. In appendix~\ref{app:lo} we give the expressions of LO singular terms. In the small $q_T$ region fixed-order expansion breaks down because of large logarithms of $\log(Q/q_T)$, and SCET can reproduce this singular behavior. 

For the large $q_T$ region, we need to include power corrections from fixed-order calculations. 
However, near $q_T \sim 30~{\rm GeV}$ where the $p^J_T>  p^{\text{min}}_{T} = 30$ GeV selection is imposed, the LO result has an artificial kink structure. The kink structure comes from the negligence of two jet events with $p_T^J<30~$GeV due to such a kinematic cut. Explicitly, at LO $p_T$ and $q_T$ are the transverse momenta of leading and subleading jets, respectively. When $q_T>30~$GeV, the lower limit of the $p_T$ integral is $q_T$. On the other hand, for $q_T<30~$GeV the lower limit is frozen at $30~$GeV. Hence, we observe such kink structure near $q_T\sim 30~$GeV. The investigation of the kink and its treatment is beyond the scope of this paper and left for future work. 
 
We also compare our theoretical calculation of the azimuthal angle decorrelation $\Delta\phi$ between the boson and the leading jet with the experimental result at $\sqrt{s}=7~$TeV in \cite{Chatrchyan:2013tna}. In the numerical integration, we boost the tree-level partonic event such that the boson and the leading jet have total transverse momentum $\vec q_T$ as
\begin{align}
    \vec q_T = q_T(\sin\phi_q, \cos\phi_q). 
\end{align}
After performing this transformation, the $Z$ boson and the leading jet are not back to back in the transverse plane. Hence, we obtain the distribution of the azimuthal angle $\Delta\phi(Z,j_1)$ between them. The comparison between NLL results and data is shown in the right plot of figure \ref{Fig:qTCMS}. The same kinematic cuts as in the experiment are imposed:
\begin{align}
&p_{T}^J>50~{\rm GeV},~~|\eta_J|<2.5,~~R=0.5, \notag \\
&p_{T}^l>20~{\rm GeV},~~|\eta_l|<2.4,~~71~{\rm GeV}<m_{ll}<111~{\rm GeV},~~p^Z_{T}>150~{\rm GeV}.
\end{align}
In principle, one needs to perform a matching between the resummed result and the fixed-order calculation to calculate the azimuthal angle decorrelation (see, e.g., \cite{Sun:2014gfa,Chen:2018fqu}). However, as we show in figure \ref{Fig:dsigmadqT}, at high $p_T$ our resummed result gives a good description even up to $q_T\sim p_T$. We then use it to calculate the normalized distribution $d\sigma/d\Delta\phi$ by integrating out $q_T$ from 0 to $150$~GeV. We find reasonable agreement with the experimental result.

\section{Summary and Perspective}\label{sec:summary}

In this paper, we construct an all-order formalism in SCET for the systematic resummation of large logarithms of the form $\log (Q/q_T)$ when $q_T\ll Q$ in boson+jet production in the small $R$ limit. More precisely, the expression  (\ref{eq:dsigmaResumFinal}) resums the logarithms of $\log(Q/q_T)$, $\log R$ and non-global logarithms, at NLL accuracy. We first carried out an analysis of the leading logarithms. We find that in the the case of $p_T \lesssim m_V$, the resummation of the logarithms $\log^2(m_V/p_T)$ is missing in the literature. In the case of $p_T \gtrsim m_V$ the effect of $\log R$ resummation only comes in at NLL accuracy. At the end, we compare our theoretical predictions with \textsc{Pythia} simulations and available experimental data \cite{Sirunyan:2018cpw,Chatrchyan:2013tna}. Within theoretical uncertainties, our results are consistent with the simulations and the data.

In the present work we obtained the resummed cross section at NLL accuracy. There are several issues that we leave for future studies. First, as shown in the plots at NLL accuracy, there are relatively large uncertainties both at small-$q_T$ and large-$q_T\sim p_T$. The small-$q_T$ region shows the sensitivity to non-perturbative physics. In this case one needs to introduce non-perturbative functions to extend the $x_T$-integral into the non-perturbative regime, and to regularize the singularity in the integration over $\phi_x$.  On the other hand, the large uncertainties at large-$q_T$ is a signature of the breaking-down of the scale separation in  (\ref{eq:scalings}). The improvement in this region is usually obtained only after matching the resummed result with a fixed-order calculation at  higher-order in $\alpha_s$. In detailed phenomenological studies, one needs to include all these improvements and possibly perform the resummation at NNLL accuracy in order to reduce the overall theoretical uncertainties. Second, the possible breaking of the transverse momentum factorization in the processes studied in this paper is another intriguing issue. On the other hand, in \textsc{Pythia} simulations we see only small non-perturbative corrections. This suggests that the $q_T$ distribution in boson+jet production can be a clean and useful probe of factorization violation and Glauber contributions. Last, but not least, the observables studied in this paper can be used to measure the jet-quenching parameters in high-energy nuclear collisions \cite{Mueller:2016gko, Mueller:2016xoc,Chen:2018fqu}. The formalism presented in this paper can be used as a unified formalism to study boson-jet correlation in both proton-proton and high-energy nuclear collisions.

\section*{Acknowledgments}

The authors thank Thomas Becher, Markus Ebert, Zhongbo Kang, Xiaohui Liu and Pier Francesco Monni for helpful discussions and comments, and Alexander Huss for sending us results for the fixed-order calculations using \textsc{NNLOjet} framework. 
This work was supported by the U.S. Department of Energy,
Office of Nuclear Physics, from DE-SC0011090. Y.-T.C. was supported in part by the LHC Theory Initiative Postdoctoral Fellowship under the National Science Foundation grant PHY-1419008.

\appendix

\section{The Hard Function at LO}
\label{app:hard}
In this appendix, we document the amplitudes and the electroweak parameters that enter the hard function in (\ref{eq:hard}) at LO. At this order, the partonic processes for a vector bonson produced associated with a jet include the following channels
\begin{align}
q(p_1) + \bar q (p_2) \to V(p_V)+g(p_J) , ~~~ q(p_1) + g (p_2) \to  V(p_V)+q(p_J),
\end{align}
where $p_i=\xi_i P_i$ with $P_i$ the proton momenta and $\xi_i$ the longitudinal momentum fractions. In these processes the partonic Mandelstam variables are defined as
\begin{align}
\hat s \equiv (p_1 + p_2)^2,~~~~\hat t \equiv (p_1 - p_V)^2,~~~~\hat u \equiv (p_2 - p_V)^2.
\end{align}
From the conservation of the $+$ and $-$ components of the momenta in the $n_1$ and $\bar {n}_1$ basis, one has
\begin{align}
\xi_1 = \frac{p_T}{\sqrt{s}} \big( e^{\eta_J} + e^{y_V} \beta_V \big),~~~\xi_2 = \frac{p_T}{\sqrt{s}} \big( e^{-\eta_J} + e^{-y_V} \beta_V \big)~~~\text{with}~~~ \beta_V = \sqrt{1 + \frac{m_V^2}{p_T^2}}. 
\end{align}
The amplitudes squared, averaged and summed over the color and spin indices in initial and final states are given by
\begin{align}\label{eq:bornmsq}
&|\overline{M}(q\bar q\to Vg)|^2=\frac{16\pi^2 \alpha_s  \alpha_{em} e_q^2(N_c^2-1)}{N_c^2}\frac{\hat t^2+\hat{u}^2+2 \hat{s}\,m_V^2}{\hat t \hat u},\notag\\
&|\overline{M}(qg\to Vq)|^2=-\frac{16\pi^2 \alpha_s  \alpha_{em} e_q^2}{N_c}\frac{\hat s^2+\hat{t}^2+2 \hat{u}\, m_V^2}{\hat s \hat t},
\end{align}
where $e_q$ is the electric charge of the quarks in the case of photon production. For $Z$ production we need to replace $e_q$ by
\begin{align}
e _ { q } ^ { 2 } \rightarrow \frac { \left( 1 - 2 \left| e _ { q } \right| \sin ^ { 2 } \theta _ { W } \right) ^ { 2 } + 4 e _ { q } ^ { 2 } \sin ^ { 4 } \theta _ { W } } { 8 \sin ^ { 2 } \theta _ { W } \cos ^ { 2 } \theta _ { W } }
\end{align}
with $\theta_W$ the weak mixing angle. In our numerical calculation the electroweak parameters we adopted are 
\begin{align}
    \alpha_{em} = 1/132.34,~~~\cos\theta_W = 0.88168,~~~m_Z = 91.1876~{\rm GeV}
\end{align}

\section{Anomalous Dimensions}
\label{app:ad}

In dimensional regularization with $d=4-2\varepsilon$, the bare strong coupling constant is replaced by the renormalized coupling constant via the relation
\begin{align}
\alpha_s^0\equiv\frac{g_{s,0}^2}{4\pi}=Z_\alpha\alpha_s(\mu)\left(\frac{\mu^2e^{\gamma_E}}{4\pi}\right)^\varepsilon.
\end{align}
An anomalous dimension is calculated from the corresponding renormalization constant $Z(\mu,\epsilon)$ according to
\begin{align}\label{eq:Gamma}
\Gamma=-\lim\limits_{\epsilon \to 0}Z^{-1}(\mu,\epsilon)\frac{d}{d\log\mu} Z(\mu,\epsilon).
\end{align}
We collect all the relevant anomalous dimensions for the NLL resummation. The running coupling constant in the $\overline{\text{MS}}$ scheme is given by the solution of
\begin{align}
\frac{d\alpha_s(\mu)}{d\log\mu}=-2 \varepsilon \alpha_s+\beta(\alpha_s(\mu)),~~~\beta(\alpha_s) = -2\alpha_s \sum_{n=0}^{\infty}\beta_n \left( \frac{\alpha_s}{4\pi} \right)^{n+1}
\end{align}
with
\begin{align}
\beta_0\equiv \frac{11}{3}C_A-\frac{2}{3}n_f,\qquad\beta_1=\frac{34}{3} C_A^2-\frac{10}{3}C_A n_f-2 C_F n_f.
\end{align}
The anomalous dimensions are expanded as a series of $\alpha_s/(4\pi)$. The cusp anomalous dimension is
\begin{align}
  \gamma_{\text{cusp}}=\frac{\alpha_s}{4\pi}\gamma_0^\text{cusp}+\left(\frac{\alpha_s}{4\pi}\right)^2\gamma_1^\text{cusp}+O(\alpha_s^3)
\end{align}
with
\begin{align}
  \gamma_0^\text{cusp}=4,\qquad \gamma_1^\text{cusp}=\left(\frac{268}{9}-\frac{4\pi^2}{3}\right)C_A-\frac{40}{9}C_F n_f,
\end{align}
and the one-loop non-cusp anomalous dimensions of jet and beam functions are 

\begin{align}
 \gamma_0^q=-3 C_F,\qquad\gamma_0^g=-\beta_0.
\end{align}

All the anomalous dimensions except $\hat{\Gamma}$ used in  (\ref{eq:dsigmaResumFinal}) consist of a cusp part, which gives the leading logarithms, and a non-cusp part, which only contributes to sub-leading logarithms. That is, these anomalous dimensions take the form
\begin{align}
\Gamma(\alpha_s)=C_\Gamma \gamma_\text{cusp}(\alpha_s)\ln\frac{Q_\Gamma^2}{\mu^2}+\gamma (\alpha_s).
\end{align}
The corresponding RG running boils down to the evaluation of the following two functions:
\begin{align}
&S ( \nu , \mu ) =  \int_{\nu}^\mu \frac{d\bar \mu}{\bar \mu} \ln \frac{\nu}{\bar \mu} \gamma_{\rm cusp}(\alpha_s(\bar \mu)),\qquad A_\gamma(\nu,\mu) = - \int_{\nu}^\mu \frac{d\bar \mu}{\bar \mu} \gamma(\alpha_s(\bar \mu)).
\end{align}
In terms of these two functions, the exponential functions in (\ref{eq:dsigmaResumFinal}) are given by
\begin{align}\label{eq:exponential}
e^{\int_{\nu}^\mu \frac{d\bar \mu}{\bar \mu}\Gamma(\bar \mu)}=\left(\frac{Q_\Gamma^2}{\nu^2}\right)^{-C_\Gamma A_{\gamma_\text{cusp}}(\nu,\mu)}e^{2 C_\Gamma S(\nu, \mu)-A_{\gamma}(\nu, \mu) }.
\end{align}
For NLL resummation, the function $S \left( \nu  , \mu \right)$ and $A _ { \gamma_\text{cusp} } \left( \nu , \mu \right) $ are given explicitly as follows,
\begin{align} \label{eq:SA}
S \left( \nu  , \mu \right) =& \frac { \gamma^\text{cusp} _ { 0 } } { 4 \beta _ { 0 } ^ { 2 } } \left\{ \frac { 4 \pi } { \alpha _ { s } \left( \nu  \right) } \left( 1 - \frac { 1 } { r } - \ln r \right) + \left( \frac {  \gamma^\text{cusp} _ { 1 } } {  \gamma^\text{cusp} _ { 0 }  } - \frac { \beta _ { 1 } } { \beta _ { 0 } } \right) ( 1 - r + \ln r )\right.\notag\\
&\left. + \frac { \beta _ { 1 } } { 2 \beta _ { 0 } } \ln ^ { 2 } r \right\},\qquad
 A _ { \gamma_\text{cusp} } \left( \nu , \mu \right) = \frac {  \gamma^\text{cusp} _ { 0 } } { 2 \beta _ { 0 } } \ln r \end{align}
with $r = \alpha _ { s } ( \mu ) / \alpha _ { s } \left( \nu \right)$. 

\section{LO singular terms}
\label{app:lo}

In this appendix we give the analytical expressions of the LO singular terms. After expanding the resummed result (\ref{eq:dsigmaNLL}) order by order in $\alpha_s$ and performing the Fourier transform, we can obtain the singular terms of the $q_T$ distribution. The LO results are given by
\begin{align}
& q_T \frac{d\sigma}{dq_T dy_V d\eta_J d^2 p_T} = \notag \\
&\sum_{ijk,ab} \mathcal{H}^{(0)}_{ij\to Vk} \int_{\xi_1}^1 \frac{d z_1}{z_1} f_{a/N_1}\left(\frac{\xi_1}{z_1},p_T\right)  \int_{\xi_2}^1 \frac{d z_2}{z_2} f_{b/N_2}\left(\frac{\xi_2}{z_2}, p_T\right) \left[ \frac{\alpha_s}{2\pi} \Sigma^{(1)}_{ij \leftarrow ab}(z_1,z_2,q_T)   \right], 
\end{align}
where the one-loop kernel $\Sigma^{(1)}_{ij \leftarrow ab}$ has the following form
\begin{align}
     \Sigma^{(1)}_{ij \leftarrow ab } = A_{ij}\delta_{ia}\delta_{jb} \delta(1-z_1)\delta(1-z_2) + \frac{1}{2}\delta(1-z_1) \delta_{ia}\mathcal{P}^{(1)}_{j \leftarrow b}(z_2) + \frac{1}{2}\delta(1-z_2) \delta_{jb}\mathcal{P}^{(1)}_{i \leftarrow a}(z_1)
\end{align}
with the coefficients 
\begin{align}
    A_{q\bar q} &= C_F \left(  4 \log\frac{\hat s}{q_T^2} - 6   \right) - 4 C_A \log R, \notag \\
    A_{qg} &= C_F \left( 2 \log\frac{\hat s\hat u}{q_T^2 \hat t} - 4\log R - 3  \right) + 2\, C_A \log\frac{\hat s\hat t}{q_T^2\hat u}   - \beta_0.
\end{align}
The one-loop Alatrelli-Parisi splitting functions are given as follows,
\begin{align}
\mathcal P^{(1)}_{q \leftarrow q}(z) &= 4 C_F\left[ \frac{1+z^2}{(1-z)_+} + \frac{3}{2}\delta(1-z) \right], \notag \\
\mathcal P^{(1)}_{g \leftarrow g}(z) &= 8 C_A \left[ \frac{z}{(1-z)_+} + \frac{1-z}{z} + z(1-z) \right] + 2\beta_0\delta(1-z), \notag \\
\mathcal { P } _ { q \leftarrow g } ^ { ( 1 ) } ( z ) &= 4 T _ { F } \left[ z ^ { 2 } + ( 1 - z ) ^ { 2 } \right], ~~~~
\mathcal { P } _ { g \leftarrow q } ^ { ( 1 ) } ( z ) = 4 C _ { F } \frac { 1 + ( 1 - z ) ^ { 2 } } { z }.
\end{align}

\bibliographystyle{JHEP}
\bibliography{jet}

\providecommand{\href}[2]{#2}\begingroup\raggedright\begin{thebibliography}{10}

\bibitem{Collins:1984kg}
J.~C. Collins, D.~E. Soper and G.~F. Sterman, \emph{{Transverse Momentum
  Distribution in Drell-Yan Pair and W and Z Boson Production}},
  \href{http://dx.doi.org/10.1016/0550-3213(85)90479-1}{\emph{Nucl. Phys.} {\bf
  B250} (1985) 199--224}.

\bibitem{Bauer:2000yr}
C.~W. Bauer, S.~Fleming, D.~Pirjol and I.~W. Stewart, \emph{{An Effective field
  theory for collinear and soft gluons: Heavy to light decays}},
  \href{http://dx.doi.org/10.1103/PhysRevD.63.114020}{\emph{Phys. Rev.} {\bf
  D63} (2001) 114020}, [\href{http://arxiv.org/abs/hep-ph/0011336}{{\tt
  hep-ph/0011336}}].

\bibitem{Bauer:2001yt}
C.~W. Bauer, D.~Pirjol and I.~W. Stewart, \emph{{Soft collinear factorization
  in effective field theory}},
  \href{http://dx.doi.org/10.1103/PhysRevD.65.054022}{\emph{Phys. Rev.} {\bf
  D65} (2002) 054022}, [\href{http://arxiv.org/abs/hep-ph/0109045}{{\tt
  hep-ph/0109045}}].

\bibitem{Bauer:2002nz}
C.~W. Bauer, S.~Fleming, D.~Pirjol, I.~Z. Rothstein and I.~W. Stewart,
  \emph{{Hard scattering factorization from effective field theory}},
  \href{http://dx.doi.org/10.1103/PhysRevD.66.014017}{\emph{Phys. Rev.} {\bf
  D66} (2002) 014017}, [\href{http://arxiv.org/abs/hep-ph/0202088}{{\tt
  hep-ph/0202088}}].

\bibitem{Beneke:2002ph}
M.~Beneke, A.~P. Chapovsky, M.~Diehl and T.~Feldmann, \emph{{Soft collinear
  effective theory and heavy to light currents beyond leading power}},
  \href{http://dx.doi.org/10.1016/S0550-3213(02)00687-9}{\emph{Nucl. Phys.}
  {\bf B643} (2002) 431--476}, [\href{http://arxiv.org/abs/hep-ph/0206152}{{\tt
  hep-ph/0206152}}].

\bibitem{Becher:2014oda}
T.~Becher, A.~Broggio and A.~Ferroglia, \emph{{Introduction to Soft-Collinear
  Effective Theory}},
  \href{http://dx.doi.org/10.1007/978-3-319-14848-9}{\emph{Lect. Notes Phys.}
  {\bf 896} (2015) pp.1--206}, [\href{http://arxiv.org/abs/1410.1892}{{\tt
  1410.1892}}].

\bibitem{Gao:2005iu}
Y.~Gao, C.~S. Li and J.~J. Liu, \emph{{Transverse momentum resummation for
  Higgs production in soft-collinear effective theory}},
  \href{http://dx.doi.org/10.1103/PhysRevD.72.114020}{\emph{Phys. Rev.} {\bf
  D72} (2005) 114020}, [\href{http://arxiv.org/abs/hep-ph/0501229}{{\tt
  hep-ph/0501229}}].

\bibitem{Mantry:2009qz}
S.~Mantry and F.~Petriello, \emph{{Factorization and Resummation of Higgs Boson
  Differential Distributions in Soft-Collinear Effective Theory}},
  \href{http://dx.doi.org/10.1103/PhysRevD.81.093007}{\emph{Phys. Rev.} {\bf
  D81} (2010) 093007}, [\href{http://arxiv.org/abs/0911.4135}{{\tt
  0911.4135}}].

\bibitem{Becher:2010tm}
T.~Becher and M.~Neubert, \emph{{{Drell-Yan} Production at Small $q_T$,
  Transverse Parton Distributions and the Collinear Anomaly}},
  \href{http://dx.doi.org/10.1140/epjc/s10052-011-1665-7}{\emph{Eur. Phys. J.}
  {\bf C71} (2011) 1665}, [\href{http://arxiv.org/abs/1007.4005}{{\tt
  1007.4005}}].

\bibitem{Becher:2011xn}
T.~Becher, M.~Neubert and D.~Wilhelm, \emph{{Electroweak Gauge-Boson Production
  at Small $q_T$: Infrared Safety from the Collinear Anomaly}},
  \href{http://dx.doi.org/10.1007/JHEP02(2012)124}{\emph{JHEP} {\bf 02} (2012)
  124}, [\href{http://arxiv.org/abs/1109.6027}{{\tt 1109.6027}}].

\bibitem{Neill:2015roa}
D.~Neill, I.~Z. Rothstein and V.~Vaidya, \emph{{The Higgs Transverse Momentum
  Distribution at NNLL and its Theoretical Errors}},
  \href{http://dx.doi.org/10.1007/JHEP12(2015)097}{\emph{JHEP} {\bf 12} (2015)
  097}, [\href{http://arxiv.org/abs/1503.00005}{{\tt 1503.00005}}].

\bibitem{Ebert:2016gcn}
M.~A. Ebert and F.~J. Tackmann, \emph{{Resummation of Transverse Momentum
  Distributions in Distribution Space}},
  \href{http://dx.doi.org/10.1007/JHEP02(2017)110}{\emph{JHEP} {\bf 02} (2017)
  110}, [\href{http://arxiv.org/abs/1611.08610}{{\tt 1611.08610}}].

\bibitem{Scimemi:2016ffw}
I.~Scimemi and A.~Vladimirov, \emph{{Power corrections and renormalons in
  Transverse Momentum Distributions}},
  \href{http://dx.doi.org/10.1007/JHEP03(2017)002}{\emph{JHEP} {\bf 03} (2017)
  002}, [\href{http://arxiv.org/abs/1609.06047}{{\tt 1609.06047}}].

\bibitem{Kang:2017cjk}
D.~Kang, C.~Lee and V.~Vaidya, \emph{{A fast and accurate method for
  perturbative resummation of transverse momentum-dependent observables}},
  \href{http://dx.doi.org/10.1007/JHEP04(2018)149}{\emph{JHEP} {\bf 04} (2018)
  149}, [\href{http://arxiv.org/abs/1710.00078}{{\tt 1710.00078}}].

\bibitem{Ebert:2018gsn}
M.~A. Ebert, I.~Moult, I.~W. Stewart, F.~J. Tackmann, G.~Vita and H.~X. Zhu,
  \emph{{Subleading power rapidity divergences and power corrections for
  q$_{T}$}}, \href{http://dx.doi.org/10.1007/JHEP04(2019)123}{\emph{JHEP} {\bf
  04} (2019) 123}, [\href{http://arxiv.org/abs/1812.08189}{{\tt 1812.08189}}].

\bibitem{Becher:2012yn}
T.~Becher, M.~Neubert and D.~Wilhelm, \emph{{Higgs-Boson Production at Small
  Transverse Momentum}},
  \href{http://dx.doi.org/10.1007/JHEP05(2013)110}{\emph{JHEP} {\bf 05} (2013)
  110}, [\href{http://arxiv.org/abs/1212.2621}{{\tt 1212.2621}}].

\bibitem{Bizon:2017rah}
W.~Bizon, P.~F. Monni, E.~Re, L.~Rottoli and P.~Torrielli,
  \emph{{Momentum-space resummation for transverse observables and the Higgs
  p$_{\perp}$ at N$^{3}$LL+NNLO}},
  \href{http://dx.doi.org/10.1007/JHEP02(2018)108}{\emph{JHEP} {\bf 02} (2018)
  108}, [\href{http://arxiv.org/abs/1705.09127}{{\tt 1705.09127}}].

\bibitem{Bizon:2018foh}
W.~Bizoń, X.~Chen, A.~Gehrmann-De~Ridder, T.~Gehrmann, N.~Glover, A.~Huss
  et~al., \emph{{Fiducial distributions in Higgs and Drell-Yan production at
  N$^{3}$LL+NNLO}},
  \href{http://dx.doi.org/10.1007/JHEP12(2018)132}{\emph{JHEP} {\bf 12} (2018)
  132}, [\href{http://arxiv.org/abs/1805.05916}{{\tt 1805.05916}}].

\bibitem{Chen:2018pzu}
X.~Chen, T.~Gehrmann, E.~W.~N. Glover, A.~Huss, Y.~Li, D.~Neill et~al.,
  \emph{{Precise QCD Description of the Higgs Boson Transverse Momentum
  Spectrum}},
  \href{http://dx.doi.org/10.1016/j.physletb.2018.11.037}{\emph{Phys. Lett.}
  {\bf B788} (2019) 425--430}, [\href{http://arxiv.org/abs/1805.00736}{{\tt
  1805.00736}}].

\bibitem{Dasgupta:2001sh}
M.~Dasgupta and G.~P. Salam, \emph{{Resummation of nonglobal QCD observables}},
  \href{http://dx.doi.org/10.1016/S0370-2693(01)00725-0}{\emph{Phys. Lett.}
  {\bf B512} (2001) 323--330}, [\href{http://arxiv.org/abs/hep-ph/0104277}{{\tt
  hep-ph/0104277}}].

\bibitem{Dasgupta:2002bw}
M.~Dasgupta and G.~P. Salam, \emph{{Accounting for coherence in interjet $E_T$
  flow: A Case study}},
  \href{http://dx.doi.org/10.1088/1126-6708/2002/03/017}{\emph{JHEP} {\bf 03}
  (2002) 017}, [\href{http://arxiv.org/abs/hep-ph/0203009}{{\tt
  hep-ph/0203009}}].

\bibitem{Sun:2014gfa}
P.~Sun, C.~P. Yuan and F.~Yuan, \emph{{Soft Gluon Resummations in Dijet
  Azimuthal Angular Correlations in Hadronic Collisions}},
  \href{http://dx.doi.org/10.1103/PhysRevLett.113.232001}{\emph{Phys. Rev.
  Lett.} {\bf 113} (2014) 232001}, [\href{http://arxiv.org/abs/1405.1105}{{\tt
  1405.1105}}].

\bibitem{Sun:2015doa}
P.~Sun, C.~P. Yuan and F.~Yuan, \emph{{Transverse Momentum Resummation for
  Dijet Correlation in Hadronic Collisions}},
  \href{http://dx.doi.org/10.1103/PhysRevD.92.094007}{\emph{Phys. Rev.} {\bf
  D92} (2015) 094007}, [\href{http://arxiv.org/abs/1506.06170}{{\tt
  1506.06170}}].

\bibitem{Chen:2018fqu}
L.~Chen, G.-Y. Qin, L.~Wang, S.-Y. Wei, B.-W. Xiao, H.-Z. Zhang et~al.,
  \emph{{Study of Isolated-photon and Jet Momentum Imbalance in $pp$ and $PbPb$
  collisions}},
  \href{http://dx.doi.org/10.1016/j.nuclphysb.2018.06.013}{\emph{Nucl. Phys.}
  {\bf B933} (2018) 306--319}, [\href{http://arxiv.org/abs/1803.10533}{{\tt
  1803.10533}}].

\bibitem{Sun:2018icb}
P.~Sun, B.~Yan, C.~P. Yuan and F.~Yuan, \emph{{Resummation of High Order
  Corrections in $Z$ Boson Plus Jet Production at the LHC}},
  \href{http://arxiv.org/abs/1810.03804}{{\tt 1810.03804}}.

\bibitem{Cao:2018ntd}
Q.-H. Cao, P.~Sun, B.~Yan, C.~P. Yuan and F.~Yuan, \emph{{Transverse Momentum
  Resummation for $t$-channel single top quark production at the LHC}},
  \href{http://dx.doi.org/10.1103/PhysRevD.98.054032}{\emph{Phys. Rev.} {\bf
  D98} (2018) 054032}, [\href{http://arxiv.org/abs/1801.09656}{{\tt
  1801.09656}}].

\bibitem{Cao:2019uor}
Q.-H. Cao, P.~Sun, B.~Yan, C.~P. Yuan and F.~Yuan, \emph{{Soft Gluon
  Resummation in $t$-channel single top quark production at the LHC}},
  \href{http://arxiv.org/abs/1902.09336}{{\tt 1902.09336}}.

\bibitem{Buffing:2018ggv}
M.~G.~A. Buffing, Z.-B. Kang, K.~Lee and X.~Liu, \emph{{A transverse momentum
  dependent framework for back-to-back photon+jet production}},
  \href{http://arxiv.org/abs/1812.07549}{{\tt 1812.07549}}.

\bibitem{Hatta:2013iba}
Y.~Hatta and T.~Ueda, \emph{{Resummation of non-global logarithms at finite
  $N_c$}}, \href{http://dx.doi.org/10.1016/j.nuclphysb.2013.06.021}{\emph{Nucl.
  Phys.} {\bf B874} (2013) 808--820},
  [\href{http://arxiv.org/abs/1304.6930}{{\tt 1304.6930}}].

\bibitem{Caron-Huot:2015bja}
S.~Caron-Huot, \emph{{Resummation of non-global logarithms and the BFKL
  equation}}, \href{http://dx.doi.org/10.1007/JHEP03(2018)036}{\emph{JHEP} {\bf
  03} (2018) 036}, [\href{http://arxiv.org/abs/1501.03754}{{\tt 1501.03754}}].

\bibitem{Larkoski:2015zka}
A.~J. Larkoski, I.~Moult and D.~Neill, \emph{{Non-Global Logarithms,
  Factorization, and the Soft Substructure of Jets}},
  \href{http://dx.doi.org/10.1007/JHEP09(2015)143}{\emph{JHEP} {\bf 09} (2015)
  143}, [\href{http://arxiv.org/abs/1501.04596}{{\tt 1501.04596}}].

\bibitem{Becher:2015hka}
T.~Becher, M.~Neubert, L.~Rothen and D.~Y. Shao, \emph{{Effective Field Theory
  for Jet Processes}},
  \href{http://dx.doi.org/10.1103/PhysRevLett.116.192001}{\emph{Phys. Rev.
  Lett.} {\bf 116} (2016) 192001}, [\href{http://arxiv.org/abs/1508.06645}{{\tt
  1508.06645}}].

\bibitem{Neill:2015nya}
D.~Neill, \emph{{The Edge of Jets and Subleading Non-Global Logs}},
  \href{http://arxiv.org/abs/1508.07568}{{\tt 1508.07568}}.

\bibitem{Becher:2016mmh}
T.~Becher, M.~Neubert, L.~Rothen and D.~Y. Shao, \emph{{Factorization and
  Resummation for Jet Processes}},
  \href{http://dx.doi.org/10.1007/JHEP11(2016)019,
  10.1007/JHEP05(2017)154}{\emph{JHEP} {\bf 11} (2016) 019},
  [\href{http://arxiv.org/abs/1605.02737}{{\tt 1605.02737}}].

\bibitem{Becher:2016omr}
T.~Becher, B.~D. Pecjak and D.~Y. Shao, \emph{{Factorization for the light-jet
  mass and hemisphere soft function}},
  \href{http://dx.doi.org/10.1007/JHEP12(2016)018}{\emph{JHEP} {\bf 12} (2016)
  018}, [\href{http://arxiv.org/abs/1610.01608}{{\tt 1610.01608}}].

\bibitem{Larkoski:2016zzc}
A.~J. Larkoski, I.~Moult and D.~Neill, \emph{{The Analytic Structure of
  Non-Global Logarithms: Convergence of the Dressed Gluon Expansion}},
  \href{http://dx.doi.org/10.1007/JHEP11(2016)089}{\emph{JHEP} {\bf 11} (2016)
  089}, [\href{http://arxiv.org/abs/1609.04011}{{\tt 1609.04011}}].

\bibitem{Neill:2016stq}
D.~Neill, \emph{{The Asymptotic Form of Non-Global Logarithms, Black Disc
  Saturation, and Gluonic Deserts}},
  \href{http://dx.doi.org/10.1007/JHEP01(2017)109}{\emph{JHEP} {\bf 01} (2017)
  109}, [\href{http://arxiv.org/abs/1610.02031}{{\tt 1610.02031}}].

\bibitem{Becher:2017nof}
T.~Becher, R.~Rahn and D.~Y. Shao, \emph{{Non-global and rapidity logarithms in
  narrow jet broadening}},
  \href{http://dx.doi.org/10.1007/JHEP10(2017)030}{\emph{JHEP} {\bf 10} (2017)
  030}, [\href{http://arxiv.org/abs/1708.04516}{{\tt 1708.04516}}].

\bibitem{Hatta:2017fwr}
Y.~Hatta, E.~Iancu, A.~H. Mueller and D.~N. Triantafyllopoulos,
  \emph{{Resumming double non-global logarithms in the evolution of a jet}},
  \href{http://dx.doi.org/10.1007/JHEP02(2018)075}{\emph{JHEP} {\bf 02} (2018)
  075}, [\href{http://arxiv.org/abs/1710.06722}{{\tt 1710.06722}}].

\bibitem{Martinez:2018ffw}
R.~Ángeles Martínez, M.~De~Angelis, J.~R. Forshaw, S.~Plätzer and M.~H.
  Seymour, \emph{{Soft gluon evolution and non-global logarithms}},
  \href{http://dx.doi.org/10.1007/JHEP05(2018)044}{\emph{JHEP} {\bf 05} (2018)
  044}, [\href{http://arxiv.org/abs/1802.08531}{{\tt 1802.08531}}].

\bibitem{Balsiger:2018ezi}
M.~Balsiger, T.~Becher and D.~Y. Shao, \emph{{Non-global logarithms in jet and
  isolation cone cross sections}},
  \href{http://dx.doi.org/10.1007/JHEP08(2018)104}{\emph{JHEP} {\bf 08} (2018)
  104}, [\href{http://arxiv.org/abs/1803.07045}{{\tt 1803.07045}}].

\bibitem{Neill:2018mmj}
D.~Neill and V.~Vaidya, \emph{{Soft evolution after a hard scattering
  process}},  \href{http://arxiv.org/abs/1803.02372}{{\tt 1803.02372}}.

\bibitem{Neill:2018yet}
D.~Neill, \emph{{Non-Global and Clustering Effects for Groomed Multi-Prong Jet
  Shapes}}, \href{http://dx.doi.org/10.1007/JHEP02(2019)114}{\emph{JHEP} {\bf
  02} (2019) 114}, [\href{http://arxiv.org/abs/1808.04897}{{\tt 1808.04897}}].

\bibitem{Balsiger:2019tne}
M.~Balsiger, T.~Becher and D.~Y. Shao, \emph{{NLL${'}$ resummation of jet
  mass}}, \href{http://dx.doi.org/10.1007/JHEP04(2019)020}{\emph{JHEP} {\bf 04}
  (2019) 020}, [\href{http://arxiv.org/abs/1901.09038}{{\tt 1901.09038}}].

\bibitem{Baier:1996sk}
R.~Baier, Y.~L. Dokshitzer, A.~H. Mueller, S.~Peigne and D.~Schiff,
  \emph{{Radiative energy loss and $p_T$ broadening of high-energy partons in
  nuclei}}, \href{http://dx.doi.org/10.1016/S0550-3213(96)00581-0}{\emph{Nucl.
  Phys.} {\bf B484} (1997) 265--282},
  [\href{http://arxiv.org/abs/hep-ph/9608322}{{\tt hep-ph/9608322}}].

\bibitem{Mueller:2016gko}
A.~H. Mueller, B.~Wu, B.-W. Xiao and F.~Yuan, \emph{{Probing Transverse
  Momentum Broadening in Heavy Ion Collisions}},
  \href{http://dx.doi.org/10.1016/j.physletb.2016.10.037}{\emph{Phys. Lett.}
  {\bf B763} (2016) 208--212}, [\href{http://arxiv.org/abs/1604.04250}{{\tt
  1604.04250}}].

\bibitem{Mueller:2016xoc}
A.~H. Mueller, B.~Wu, B.-W. Xiao and F.~Yuan, \emph{{Medium Induced Transverse
  Momentum Broadening in Hard Processes}},
  \href{http://dx.doi.org/10.1103/PhysRevD.95.034007}{\emph{Phys. Rev.} {\bf
  D95} (2017) 034007}, [\href{http://arxiv.org/abs/1608.07339}{{\tt
  1608.07339}}].

\bibitem{Dai:2012am}
W.~Dai, I.~Vitev and B.-W. Zhang, \emph{{Momentum imbalance of isolated
  photon-tagged jet production at RHIC and LHC}},
  \href{http://dx.doi.org/10.1103/PhysRevLett.110.142001}{\emph{Phys. Rev.
  Lett.} {\bf 110} (2013) 142001}, [\href{http://arxiv.org/abs/1207.5177}{{\tt
  1207.5177}}].

\bibitem{Neufeld:2012df}
R.~B. Neufeld and I.~Vitev, \emph{{The $Z^0$-tagged jet event asymmetry in
  heavy-ion collisions at the CERN Large Hadron Collider}},
  \href{http://dx.doi.org/10.1103/PhysRevLett.108.242001}{\emph{Phys. Rev.
  Lett.} {\bf 108} (2012) 242001}, [\href{http://arxiv.org/abs/1202.5556}{{\tt
  1202.5556}}].

\bibitem{Wang:2013cia}
X.-N. Wang and Y.~Zhu, \emph{{Medium Modification of $\gamma$-jets in
  High-energy Heavy-ion Collisions}},
  \href{http://dx.doi.org/10.1103/PhysRevLett.111.062301}{\emph{Phys. Rev.
  Lett.} {\bf 111} (2013) 062301}, [\href{http://arxiv.org/abs/1302.5874}{{\tt
  1302.5874}}].

\bibitem{Casalderrey-Solana:2015vaa}
J.~Casalderrey-Solana, D.~C. Gulhan, J.~G. Milhano, D.~Pablos and K.~Rajagopal,
  \emph{{Predictions for Boson-Jet Observables and Fragmentation Function
  Ratios from a Hybrid Strong/Weak Coupling Model for Jet Quenching}},
  \href{http://dx.doi.org/10.1007/JHEP03(2016)053}{\emph{JHEP} {\bf 03} (2016)
  053}, [\href{http://arxiv.org/abs/1508.00815}{{\tt 1508.00815}}].

\bibitem{Chien:2015hda}
Y.-T. Chien and I.~Vitev, \emph{{Towards the understanding of jet shapes and
  cross sections in heavy ion collisions using soft-collinear effective
  theory}}, \href{http://dx.doi.org/10.1007/JHEP05(2016)023}{\emph{JHEP} {\bf
  05} (2016) 023}, [\href{http://arxiv.org/abs/1509.07257}{{\tt 1509.07257}}].

\bibitem{KunnawalkamElayavalli:2016ttl}
R.~Kunnawalkam~Elayavalli and K.~C. Zapp, \emph{{Simulating V+jet processes in
  heavy ion collisions with JEWEL}},
  \href{http://dx.doi.org/10.1140/epjc/s10052-016-4534-6}{\emph{Eur. Phys. J.}
  {\bf C76} (2016) 695}, [\href{http://arxiv.org/abs/1608.03099}{{\tt
  1608.03099}}].

\bibitem{Kang:2017xnc}
Z.-B. Kang, I.~Vitev and H.~Xing, \emph{{Vector-boson-tagged jet production in
  heavy ion collisions at energies available at the CERN Large Hadron
  Collider}}, \href{http://dx.doi.org/10.1103/PhysRevC.96.014912}{\emph{Phys.
  Rev.} {\bf C96} (2017) 014912}, [\href{http://arxiv.org/abs/1702.07276}{{\tt
  1702.07276}}].

\bibitem{Aad:2012en}
{\scshape ATLAS} collaboration, G.~Aad et~al., \emph{{Study of jets produced in
  association with a $W$ boson in $pp$ collisions at $\sqrt{s}=7$ TeV with the
  ATLAS detector}},
  \href{http://dx.doi.org/10.1103/PhysRevD.85.092002}{\emph{Phys. Rev.} {\bf
  D85} (2012) 092002}, [\href{http://arxiv.org/abs/1201.1276}{{\tt
  1201.1276}}].

\bibitem{Chatrchyan:2013tna}
{\scshape CMS} collaboration, S.~Chatrchyan et~al., \emph{{Event Shapes and
  Azimuthal Correlations in $Z$ + Jets Events in $pp$ Collisions at
  $\sqrt{s}=7$ TeV}},
  \href{http://dx.doi.org/10.1016/j.physletb.2013.04.025}{\emph{Phys. Lett.}
  {\bf B722} (2013) 238--261}, [\href{http://arxiv.org/abs/1301.1646}{{\tt
  1301.1646}}].

\bibitem{Khachatryan:2016crw}
{\scshape CMS} collaboration, V.~Khachatryan et~al., \emph{{Measurements of
  differential production cross sections for a Z boson in association with jets
  in pp collisions at $ \sqrt{s}=8 $ TeV}},
  \href{http://dx.doi.org/10.1007/JHEP04(2017)022}{\emph{JHEP} {\bf 04} (2017)
  022}, [\href{http://arxiv.org/abs/1611.03844}{{\tt 1611.03844}}].

\bibitem{Sirunyan:2017jic}
{\scshape CMS} collaboration, A.~M. Sirunyan et~al., \emph{{Study of Jet
  Quenching with $Z+\text{jet}$ Correlations in Pb-Pb and $pp$ Collisions at
  ${\sqrt{s}_{NN}}=5.02\text{ }\text{ }\mathrm{TeV}$}},
  \href{http://dx.doi.org/10.1103/PhysRevLett.119.082301}{\emph{Phys. Rev.
  Lett.} {\bf 119} (2017) 082301}, [\href{http://arxiv.org/abs/1702.01060}{{\tt
  1702.01060}}].

\bibitem{Sirunyan:2018cpw}
{\scshape CMS} collaboration, A.~M. Sirunyan et~al., \emph{{Measurement of
  differential cross sections for Z boson production in association with jets
  in proton-proton collisions at $\sqrt{s} = 13$ TeV}},
  \href{http://dx.doi.org/10.1140/epjc/s10052-018-6373-0}{\emph{Eur. Phys. J.}
  {\bf C78} (2018) 965}, [\href{http://arxiv.org/abs/1804.05252}{{\tt
  1804.05252}}].

\bibitem{Aaboud:2017kff}
{\scshape ATLAS} collaboration, M.~Aaboud et~al., \emph{{Measurement of the
  cross section for isolated-photon plus jet production in $pp$ collisions at
  $\sqrt s=13$ TeV using the ATLAS detector}},
  \href{http://dx.doi.org/10.1016/j.physletb.2018.03.035}{\emph{Phys. Lett.}
  {\bf B780} (2018) 578--602}, [\href{http://arxiv.org/abs/1801.00112}{{\tt
  1801.00112}}].

\bibitem{Chien:2015cka}
Y.-T. Chien, A.~Hornig and C.~Lee, \emph{{Soft-collinear mode for jet cross
  sections in soft collinear effective theory}},
  \href{http://dx.doi.org/10.1103/PhysRevD.93.014033}{\emph{Phys. Rev.} {\bf
  D93} (2016) 014033}, [\href{http://arxiv.org/abs/1509.04287}{{\tt
  1509.04287}}].

\bibitem{Kolodrubetz:2016dzb}
D.~W. Kolodrubetz, P.~Pietrulewicz, I.~W. Stewart, F.~J. Tackmann and W.~J.
  Waalewijn, \emph{{Factorization for Jet Radius Logarithms in Jet Mass Spectra
  at the LHC}}, \href{http://dx.doi.org/10.1007/JHEP12(2016)054}{\emph{JHEP}
  {\bf 12} (2016) 054}, [\href{http://arxiv.org/abs/1605.08038}{{\tt
  1605.08038}}].

\bibitem{Collins:2007nk}
J.~Collins and J.-W. Qiu, \emph{{$k_{T}$ factorization is violated in
  production of high-transverse-momentum particles in hadron-hadron
  collisions}}, \href{http://dx.doi.org/10.1103/PhysRevD.75.114014}{\emph{Phys.
  Rev.} {\bf D75} (2007) 114014}, [\href{http://arxiv.org/abs/0705.2141}{{\tt
  0705.2141}}].

\bibitem{Rogers:2010dm}
T.~C. Rogers and P.~J. Mulders, \emph{{No Generalized TMD-Factorization in
  Hadro-Production of High Transverse Momentum Hadrons}},
  \href{http://dx.doi.org/10.1103/PhysRevD.81.094006}{\emph{Phys. Rev.} {\bf
  D81} (2010) 094006}, [\href{http://arxiv.org/abs/1001.2977}{{\tt
  1001.2977}}].

\bibitem{Catani:2011st}
S.~Catani, D.~de~Florian and G.~Rodrigo, \emph{{Space-like (versus time-like)
  collinear limits in QCD: Is factorization violated?}},
  \href{http://dx.doi.org/10.1007/JHEP07(2012)026}{\emph{JHEP} {\bf 07} (2012)
  026}, [\href{http://arxiv.org/abs/1112.4405}{{\tt 1112.4405}}].

\bibitem{Forshaw:2012bi}
J.~R. Forshaw, M.~H. Seymour and A.~Siodmok, \emph{{On the Breaking of
  Collinear Factorization in QCD}},
  \href{http://dx.doi.org/10.1007/JHEP11(2012)066}{\emph{JHEP} {\bf 11} (2012)
  066}, [\href{http://arxiv.org/abs/1206.6363}{{\tt 1206.6363}}].

\bibitem{Rothstein:2016bsq}
I.~Z. Rothstein and I.~W. Stewart, \emph{{An Effective Field Theory for Forward
  Scattering and Factorization Violation}},
  \href{http://dx.doi.org/10.1007/JHEP08(2016)025}{\emph{JHEP} {\bf 08} (2016)
  025}, [\href{http://arxiv.org/abs/1601.04695}{{\tt 1601.04695}}].

\bibitem{Hill:2002vw}
R.~J. Hill and M.~Neubert, \emph{{Spectator interactions in soft collinear
  effective theory}},
  \href{http://dx.doi.org/10.1016/S0550-3213(03)00116-0}{\emph{Nucl. Phys.}
  {\bf B657} (2003) 229--256}, [\href{http://arxiv.org/abs/hep-ph/0211018}{{\tt
  hep-ph/0211018}}].

\bibitem{Catani:2010pd}
S.~Catani and M.~Grazzini, \emph{{QCD transverse-momentum resummation in gluon
  fusion processes}},
  \href{http://dx.doi.org/10.1016/j.nuclphysb.2010.12.007}{\emph{Nucl. Phys.}
  {\bf B845} (2011) 297--323}, [\href{http://arxiv.org/abs/1011.3918}{{\tt
  1011.3918}}].

\bibitem{Collins:1981uk}
J.~C. Collins and D.~E. Soper, \emph{{Back-To-Back Jets in QCD}},
  \href{http://dx.doi.org/10.1016/0550-3213(81)90339-4}{\emph{Nucl. Phys.} {\bf
  B193} (1981) 381}.

\bibitem{Collins:1981uw}
J.~C. Collins and D.~E. Soper, \emph{{Parton Distribution and Decay
  Functions}},
  \href{http://dx.doi.org/10.1016/0550-3213(82)90021-9}{\emph{Nucl. Phys.} {\bf
  B194} (1982) 445--492}.

\bibitem{Stewart:2009yx}
I.~W. Stewart, F.~J. Tackmann and W.~J. Waalewijn, \emph{{Factorization at the
  LHC: From PDFs to Initial State Jets}},
  \href{http://dx.doi.org/10.1103/PhysRevD.81.094035}{\emph{Phys. Rev.} {\bf
  D81} (2010) 094035}, [\href{http://arxiv.org/abs/0910.0467}{{\tt
  0910.0467}}].

\bibitem{Dokshitzer:1987nm}
Y.~L. Dokshitzer, V.~A. Khoze, S.~I. Troian and A.~H. Mueller, \emph{{QCD
  Coherence in High-Energy Reactions}},
  \href{http://dx.doi.org/10.1103/RevModPhys.60.373}{\emph{Rev. Mod. Phys.}
  {\bf 60} (1988) 373}.

\bibitem{Cacciari:2008gp}
M.~Cacciari, G.~P. Salam and G.~Soyez, \emph{{The anti-$k_t$ jet clustering
  algorithm}},
  \href{http://dx.doi.org/10.1088/1126-6708/2008/04/063}{\emph{JHEP} {\bf 04}
  (2008) 063}, [\href{http://arxiv.org/abs/0802.1189}{{\tt 0802.1189}}].

\bibitem{Becher:2011dz}
T.~Becher and G.~Bell, \emph{{Analytic Regularization in Soft-Collinear
  Effective Theory}},
  \href{http://dx.doi.org/10.1016/j.physletb.2012.05.016}{\emph{Phys. Lett.}
  {\bf B713} (2012) 41--46}, [\href{http://arxiv.org/abs/1112.3907}{{\tt
  1112.3907}}].

\bibitem{Chiu:2011qc}
J.-y. Chiu, A.~Jain, D.~Neill and I.~Z. Rothstein, \emph{{The Rapidity
  Renormalization Group}},
  \href{http://dx.doi.org/10.1103/PhysRevLett.108.151601}{\emph{Phys. Rev.
  Lett.} {\bf 108} (2012) 151601}, [\href{http://arxiv.org/abs/1104.0881}{{\tt
  1104.0881}}].

\bibitem{Chiu:2012ir}
J.-Y. Chiu, A.~Jain, D.~Neill and I.~Z. Rothstein, \emph{{A Formalism for the
  Systematic Treatment of Rapidity Logarithms in Quantum Field Theory}},
  \href{http://dx.doi.org/10.1007/JHEP05(2012)084}{\emph{JHEP} {\bf 05} (2012)
  084}, [\href{http://arxiv.org/abs/1202.0814}{{\tt 1202.0814}}].

\bibitem{Becher:2015gsa}
T.~Becher and X.~Garcia~i Tormo, \emph{{Factorization and resummation for
  transverse thrust}},
  \href{http://dx.doi.org/10.1007/JHEP06(2015)071}{\emph{JHEP} {\bf 06} (2015)
  071}, [\href{http://arxiv.org/abs/1502.04136}{{\tt 1502.04136}}].

\bibitem{Becher:2011fc}
T.~Becher, C.~Lorentzen and M.~D. Schwartz, \emph{{Resummation for W and Z
  production at large pT}},
  \href{http://dx.doi.org/10.1103/PhysRevLett.108.012001}{\emph{Phys. Rev.
  Lett.} {\bf 108} (2012) 012001}, [\href{http://arxiv.org/abs/1106.4310}{{\tt
  1106.4310}}].

\bibitem{Ellis:2010rwa}
S.~D. Ellis, C.~K. Vermilion, J.~R. Walsh, A.~Hornig and C.~Lee, \emph{{Jet
  Shapes and Jet Algorithms in SCET}},
  \href{http://dx.doi.org/10.1007/JHEP11(2010)101}{\emph{JHEP} {\bf 11} (2010)
  101}, [\href{http://arxiv.org/abs/1001.0014}{{\tt 1001.0014}}].

\bibitem{Dasgupta:2012hg}
M.~Dasgupta, K.~Khelifa-Kerfa, S.~Marzani and M.~Spannowsky, \emph{{On jet mass
  distributions in Z+jet and dijet processes at the LHC}},
  \href{http://dx.doi.org/10.1007/JHEP10(2012)126}{\emph{JHEP} {\bf 10} (2012)
  126}, [\href{http://arxiv.org/abs/1207.1640}{{\tt 1207.1640}}].

\bibitem{Sjostrand:2014zea}
T.~Sjostrand, S.~Ask, J.~R. Christiansen, R.~Corke, N.~Desai, P.~Ilten et~al.,
  \emph{{An Introduction to PYTHIA 8.2}},
  \href{http://dx.doi.org/10.1016/j.cpc.2015.01.024}{\emph{Comput. Phys.
  Commun.} {\bf 191} (2015) 159--177},
  [\href{http://arxiv.org/abs/1410.3012}{{\tt 1410.3012}}].

\bibitem{Dulat:2015mca}
S.~Dulat, T.-J. Hou, J.~Gao, M.~Guzzi, J.~Huston, P.~Nadolsky et~al.,
  \emph{{New parton distribution functions from a global analysis of quantum
  chromodynamics}},
  \href{http://dx.doi.org/10.1103/PhysRevD.93.033006}{\emph{Phys. Rev.} {\bf
  D93} (2016) 033006}, [\href{http://arxiv.org/abs/1506.07443}{{\tt
  1506.07443}}].

\bibitem{Campbell:2002tg}
J.~M. Campbell and R.~K. Ellis, \emph{{Next-to-leading order corrections to $W
  + 2$ jet and $Z + 2$ jet production at hadron colliders}},
  \href{http://dx.doi.org/10.1103/PhysRevD.65.113007}{\emph{Phys. Rev.} {\bf
  D65} (2002) 113007}, [\href{http://arxiv.org/abs/hep-ph/0202176}{{\tt
  hep-ph/0202176}}].

\bibitem{Campbell:2003hd}
J.~M. Campbell, R.~K. Ellis and D.~L. Rainwater, \emph{{Next-to-leading order
  QCD predictions for $W + 2$ jet and $Z + 2$ jet production at the CERN LHC}},
  \href{http://dx.doi.org/10.1103/PhysRevD.68.094021}{\emph{Phys. Rev.} {\bf
  D68} (2003) 094021}, [\href{http://arxiv.org/abs/hep-ph/0308195}{{\tt
  hep-ph/0308195}}].

\end{thebibliography}\endgroup


\end{document}